\documentclass{ws-procs975x65}
\usepackage{amssymb} %maths
\usepackage{amsmath} %maths
\usepackage{bm}
\newcommand{\beq}{\begin{equation}}
\newcommand{\eeq}{\end{equation}}
\newcommand{\ba}{\begin{array}{ccc}}
\newcommand{\ea}{\end{array}}
\newcommand{\nn}{\nonumber \\}

\def\bea{\begin{eqnarray}}
\def\eea{\end{eqnarray}}
\begin{document}

\title{The quantum phases of matter}

\author{Subir Sachdev}
  
\address{Department of Physics, Harvard University, Cambridge MA 02138}

\begin{abstract}
I present a selective survey of the phases of quantum matter with varieties of many-particle
quantum entanglement. I classify the phases as gapped, conformal, or compressible quantum matter.
Gapped quantum matter is illustrated by a simple discussion of the $\mathbb{Z}_2$ spin liquid, and connections
are made to topological field theories. I discuss how conformal matter is realized at quantum critical points of realistic
lattice models, and make connections to a number of experimental systems. Recent progress in our understanding of
compressible quantum phases which are not Fermi liquids is summarized. Finally, I discuss how
the strongly-coupled phases of quantum matter may be described by gauge-gravity duality. The structure
of the $N_c \rightarrow \infty$ limit of SU($N_c$) gauge theory, coupled
to adjoint fermion matter at non-zero density, suggests aspects of gravitational duals of compressible quantum matter.
\end{abstract}

%\keywords{}
\bodymatter
\begin{center}
{\tt Rapporteur presentation at the 25th Solvay Conference on Physics,\\ {\em The Theory of 
the Quantum World\/}, Brussels, Oct 19-22, 2011}
\end{center}
\section{Introduction}
\label{sec:intro}

Some of the most stringest tests and profound consequences of the quantum theory
appear in its application to large numbers of electrons in crystals. Sommerfeld and Bloch's
early theory of electronic motion in metals treated the electrons as largely independent particles
moving in the periodic potential created by the crystalline background. The basic principles were
the same as in Schr\"odinger's theory of atomic structure: the electrons occupy `orbitals' obtained by solving
the single particle Schr\"odinger equation, and mainly feel each other via Pauli's exclusion principle. Extensions of this
theory have since led to a remarkably complete and quantitative 
understanding of most common metals, superconductors, and insulators.

In the past thirty years, the application of quantum theory to many particle physics has entered a new terrain.
It has become clear that many phases of quantum matter cannot be described by extensions of the one-particle
theory, and new paradigms of the quantum behavior of many particles are needed. In an influential early paper,\cite{epr}
Einstein, Podolsky, and Rosen (EPR) emphasized that the quantum theory implied non-local correlations between states
of well separated electrons which they found unpalatable. Bell later showed\cite{bell} that such non-local correlations could not
be obtained in any classical hidden variable theory. Today, it is common to refer to such non-local EPR correlations as
{\em quantum entanglement\/}. Many varieties of entanglement play a fundamental role in the structure of the
phases of quantum matter, and it is often long-ranged. Remarkably, the long-range entanglement appears
in the natural state of the many materials at low enough temperatures, and does not require delicate
preparation of specific quantum states after protection from environmental perturbations.

\begin{figure}[htbp]
\begin{center}
 \includegraphics[width=2.5in]{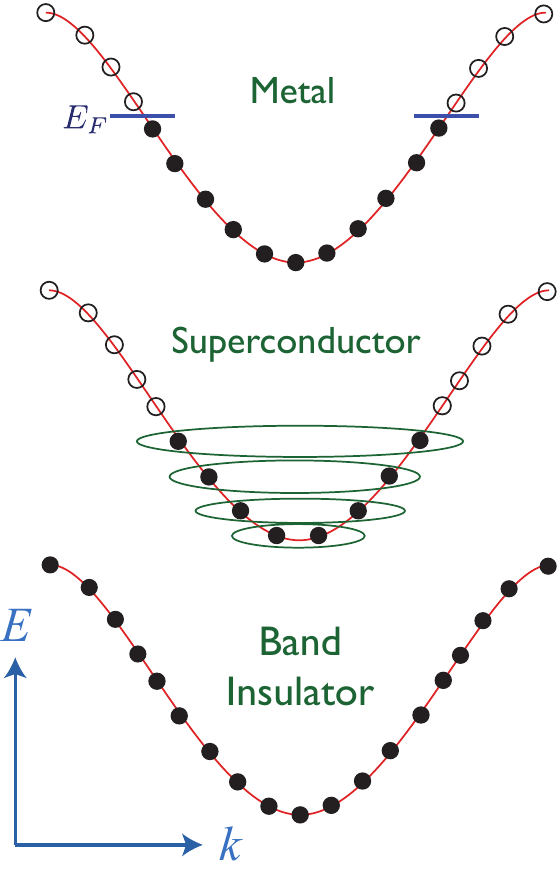}
 \caption{Schematic of the phases of matter which can be described by extensions of the independent electron theory.}
\label{fig:band}
\end{center}
\end{figure}

The structure of Sommerfeld-Bloch theory of metals is summarized in Fig.~\ref{fig:band}. 
The electrons occupy single-particle states labelled by a momentum ${\bm k}$ below the
Fermi energy $E_F$. The states with energy equal to $E_F$ define a $(d-1)$-dimensional `Fermi surface' in momentum space
(in spatial dimension $d$), and the low energy excitations across the Fermi surface are
responsible for the metallic conduction. When the Fermi energy lies in an energy gap, then the occupied states completely fill
a set of bands, and there is an energy gap to all electronic excitations: this defines a band insulator, and the band-filling criterion
requires that there be an even number of electrons per unit cell. Upon including the effect of electron-electron
interactions, which can be quite large, both the metal and the band insulator remain adiabatically
connected to the states of free electrons illustrated in Fig.~\ref{fig:band}. Finally, in the Bardeen-Cooper-Schrieffer theory, 
a superconductor is obtained when the electrons
form pairs, and the pairs Bose condense. In this case, the ground state is typically adiabatically connected to the Bose-Einstein condensate
of electron pairs, which is a simple product of single boson states.

Here, I will give a selective survey of the phases of quantum matter which cannot be adiabatically connected to free electron states, 
and which realize the different flavors of many-body quantum entanglement.
I will organize the discussion by classifying the states by the nature of their excitation spectrum.
Readers interested primarily in strange metals can skip ahead to Section~\ref{sec:nfl}.

First, in Section~\ref{sec:gap}, 
I will consider phases in which there is a gap to all excitations in the bulk matter (although, there may be gapless 
excitations along the boundary). Despite the absence of low energy excitations, such states can have subtle forms
of many body entanglement which are described by topological field theories.

Section~\ref{sec:cft} will consider states which are gapless, with the zero energy excitations
only found at isolated points in the Brillouin zone. Such states often have an excitation spectrum of massless relativistic particles,
with the role of the velocity of light being played by a smaller velocity associated with the lattice Hamiltonian. Moreover,
many such states are described by a quantum field theory which is invariant under conformal transformations of spacetime,
and hence Section~\ref{sec:cft} will describe `conformal' quantum matter.

Section~\ref{sec:nfl} will turn to `compressible' quantum matter, in which the density of particles can be varied
smoothly by an external chemical potential, without changing the basic characteristics of the phase. All known examples of such
phases have zero energy excitations along a $(d-1)$-dimensional surface in momentum space, just as in the free electron 
metal.\footnote{The categories of conformal and compressible matter overlap in $d=1$. Almost
all of our discussion will focus on $d=2$ and higher.}
However, there is much experimental and theoretical interest in describing so-called `strange metals', which are compressible
states not smoothly connected to the free electron metal. I will summarize recent theoretical studies of strange metals.

Section~\ref{sec:string} will discuss emerging connections between the above studies of the phases of quantum matter
and string theory. I will summarize a new perspective on the gravity duals of compressible states.

\newpage
\section{Gapped quantum matter}
\label{sec:gap}

The earliest discussion of the non-trivial phases of gapped quantum matter emerged
in studies of Mott insulators. These insulators appear when Coulomb repulsion is the primary impediment to the motion of electrons, 
rather than the absence of single particle states for band insulators. In a situation where there are an odd number of electrons per unit cell,
the independent electron approach necessarily leads to partially filled bands, and hence predicts the presence of a Fermi surface and metallic behavior. However, if the Coulomb repulsion, $U$, is large compared to the bandwidth, $W$, then the motion of the charge of the electrons can be sufficiently suppressed to yield a vanishing conductivity in the limit of zero temperature. 

\begin{figure}[htbp]
\begin{center}
 \includegraphics[width=3.4in]{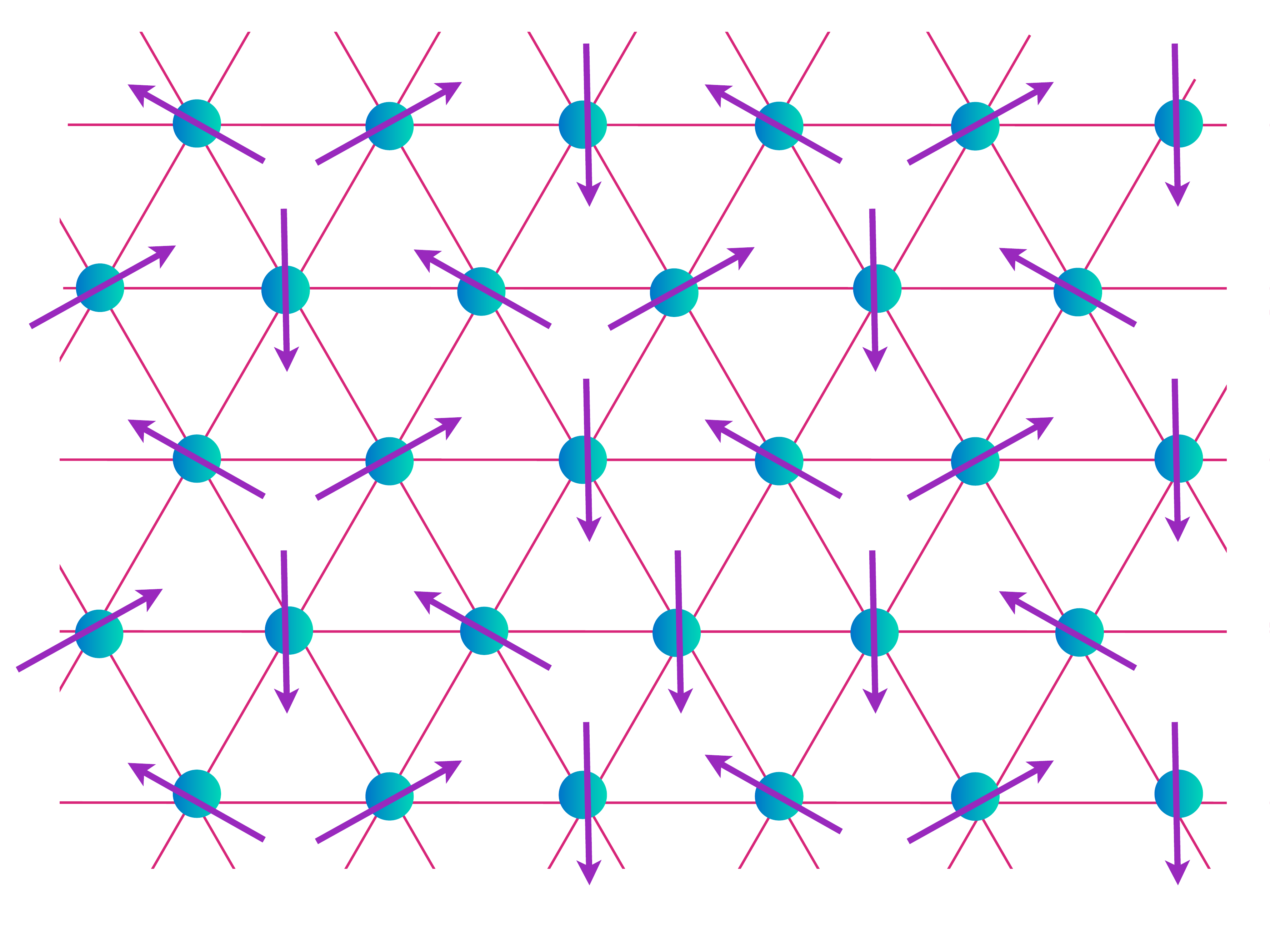}
 \caption{Ground state of the Heisenberg antiferromagnet on the triangular lattice with long-range antiferromagnetic order. This state is not
 an example of gapped quantum matter.}
\label{fig:afmtri}
\end{center}
\end{figure}

For a simple example of a Mott insulator, consider electrons hopping in a single band on the triangular lattice.
After the Coulomb repulsion localizes the electron charge, the Hilbert space can be truncated to the quantum states
in which there is precisely one electron on each site. This Hilbert space is not trivial because we have not specified the spins
of the electrons: indeed the spin degeneracy implies that there are $2^N$ states in this truncated Hilbert space,
in a lattice of $N$ sites. The degeneracy of these spin states is lifted by virtual processes involving charge fluctuations,
and these lead to the Hamiltonian of a Heisenberg antiferromagnet
\beq
\mathcal{H}_{AF} = J \sum_{\langle ij \rangle} \vec{S}_i \cdot \vec{S}_j + \ldots
\label{haf}
\eeq
where $\vec{S}_i$ is the $S=1/2$ spin operator acting on the electron on site $i$, $J \sim W^2/U > 0$ is the exchange
interaction generated by the virtual processes, and the ellipses indicates omitted terms generated at higher orders in a $W/U$ expansion.
The exact ground state of $\mathcal{H}_{AF}$ is not known. However, for the truncated Hamiltonian with only nearest neighbor exchange,
there is good numerical evidence \cite{cherny} that the ground state has long-range antiferromagnetic (N\'eel) order of the type illustrated
in Fig.~\ref{fig:afmtri}.
In this state, the spins behave in an essentially classical manner. Each spin has a mean polarization, oriented along the arrows in 
Fig.~\ref{fig:afmtri}, and there are quantum spin fluctuations about this average direction. There are gapless spin-wave excitations above
this ground state, and so this state is not an example of gapped quantum matter. Furthermore, this ground state is adiabatically connected
to the classical state with frozen spins, and so this state does not carry the kind of quantum entanglement we are seeking.

\begin{figure}[htbp]
\begin{center}
 \includegraphics[width=3.4in]{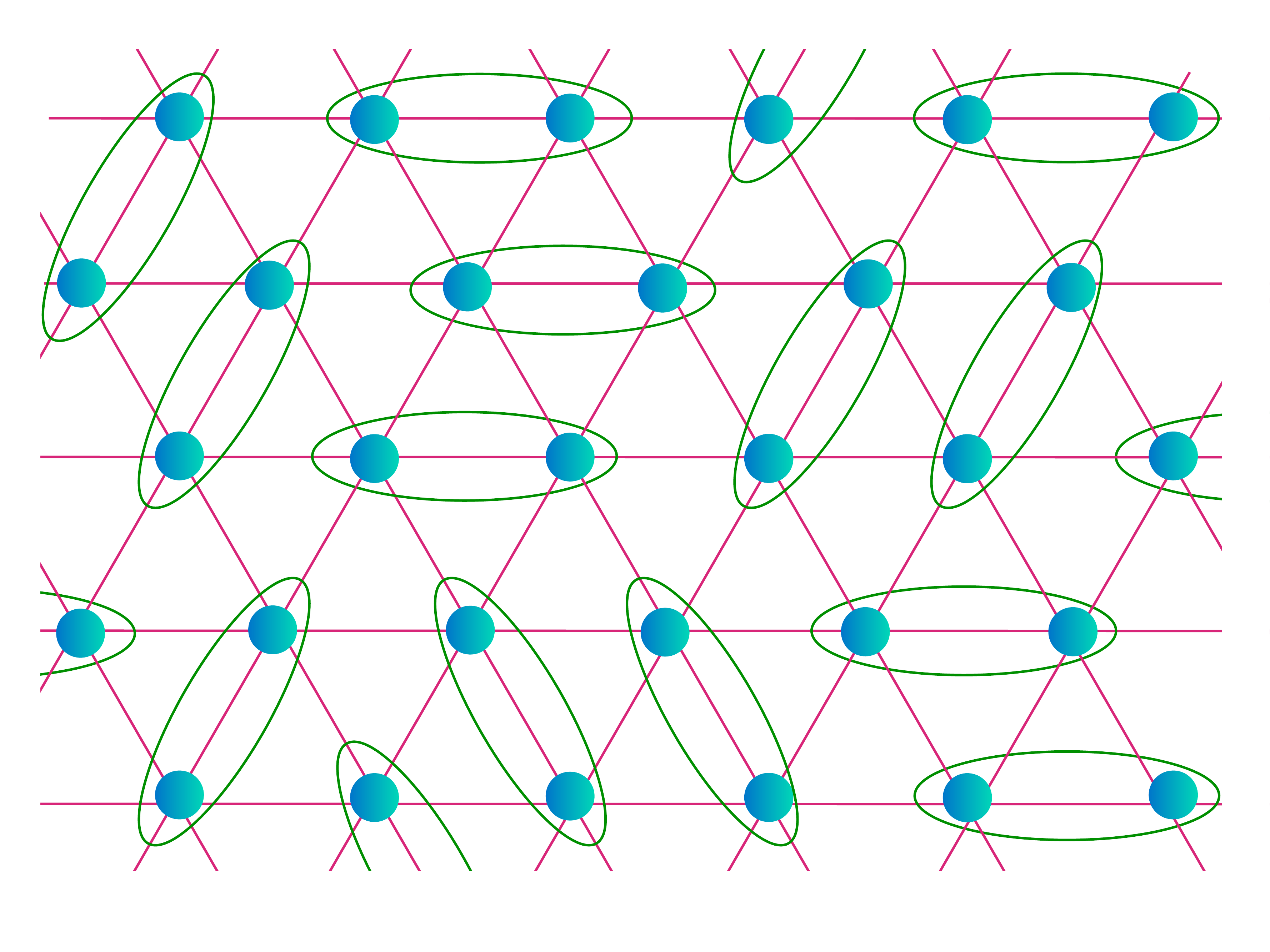}
 \caption{A snapshot of the RVB state on the triangular lattice. Each ellipse represents a singlet valence bond,
 $(| \uparrow \downarrow \rangle - | \downarrow \uparrow \rangle)/\sqrt{2}$. The RVB state is a superposition
 of all different singlet pairings, of which only one is shown above.}
\label{fig:afmrvb}
\end{center}
\end{figure}

\begin{figure}[htbp]
\begin{center}
 \includegraphics[width=3.4in]{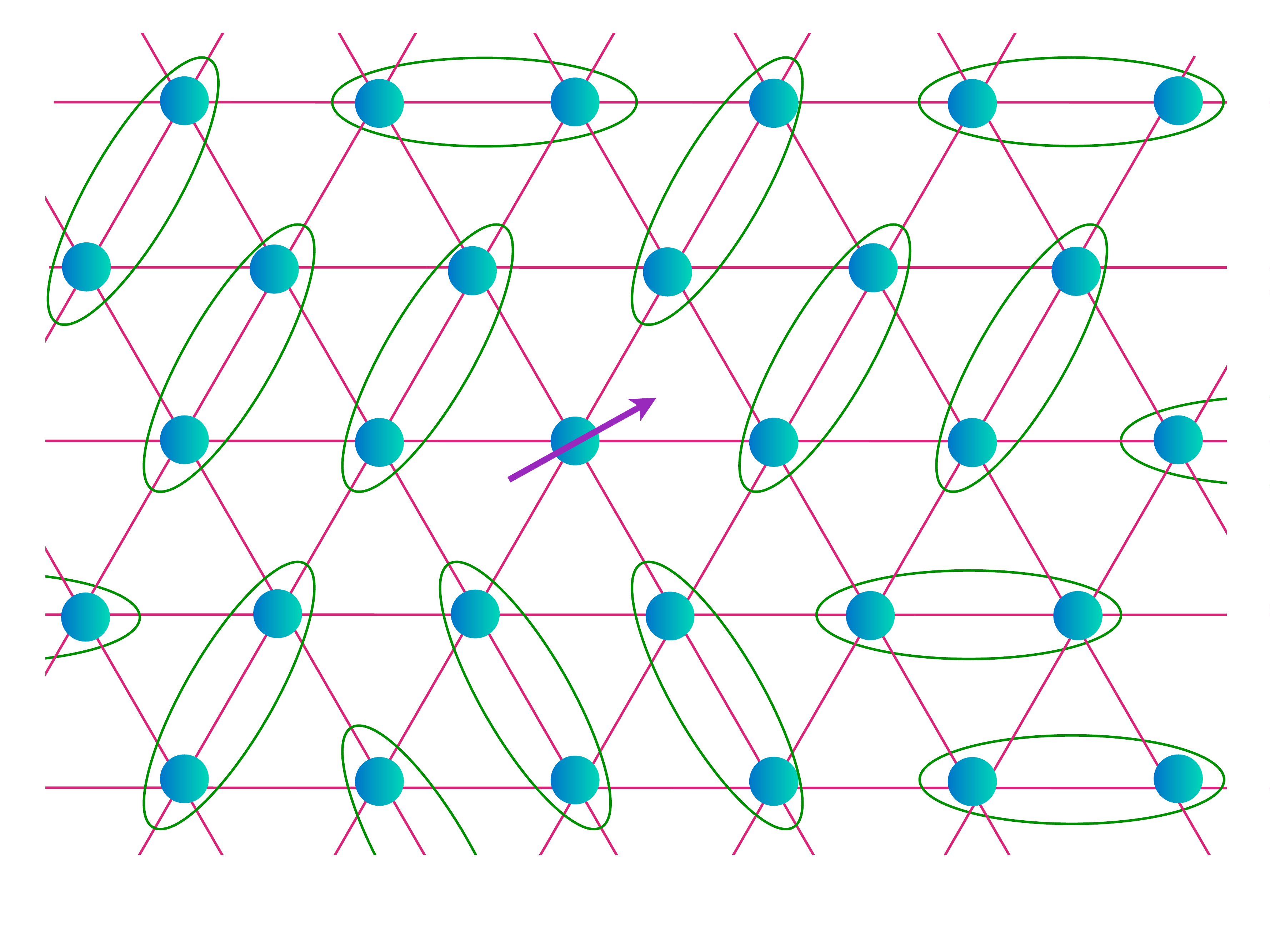}
 \caption{A spinon excitation of the RVB state. The spinon carries spin $S=1/2$ but is electrically neutral.}
\label{fig:spinon}
\end{center}
\end{figure}

To obtain gapped quantum matter, we have to consider the possibility of a different type of ground state of antiferromagnets
in the class described by $\mathcal{H}_{AF}$. This is the resonating valence bond (RVB) state of 
Fazekas and Anderson\cite{fazekas}, illustrated in Fig.~\ref{fig:afmrvb}.
This state is a linear superposition of the very large number of possible singlet pairings between the electrons.
It thus generalizes the chemical resonance of $\pi$-bonds in the benzene ring to an infinite number of electrons on a lattice.
Such a RVB wavefunction was written down early on by Pauling,\cite{pauling} who proposed it as a theory of a correlated metal.
Anderson\cite{pwa} applied the RVB state to the spin physics of a Mott insulator, and Kivelson {\em et al.}\cite{KRS} noted that it
exhibits the phenomenon of spin-charge separation: there are {\em spinon\/} excitations which carry spin $S=1/2$
but do not transfer any charge, as shown in Fig.~\ref{fig:spinon}.

\begin{figure}[htbp]
\begin{center}
 \includegraphics[width=3.4in]{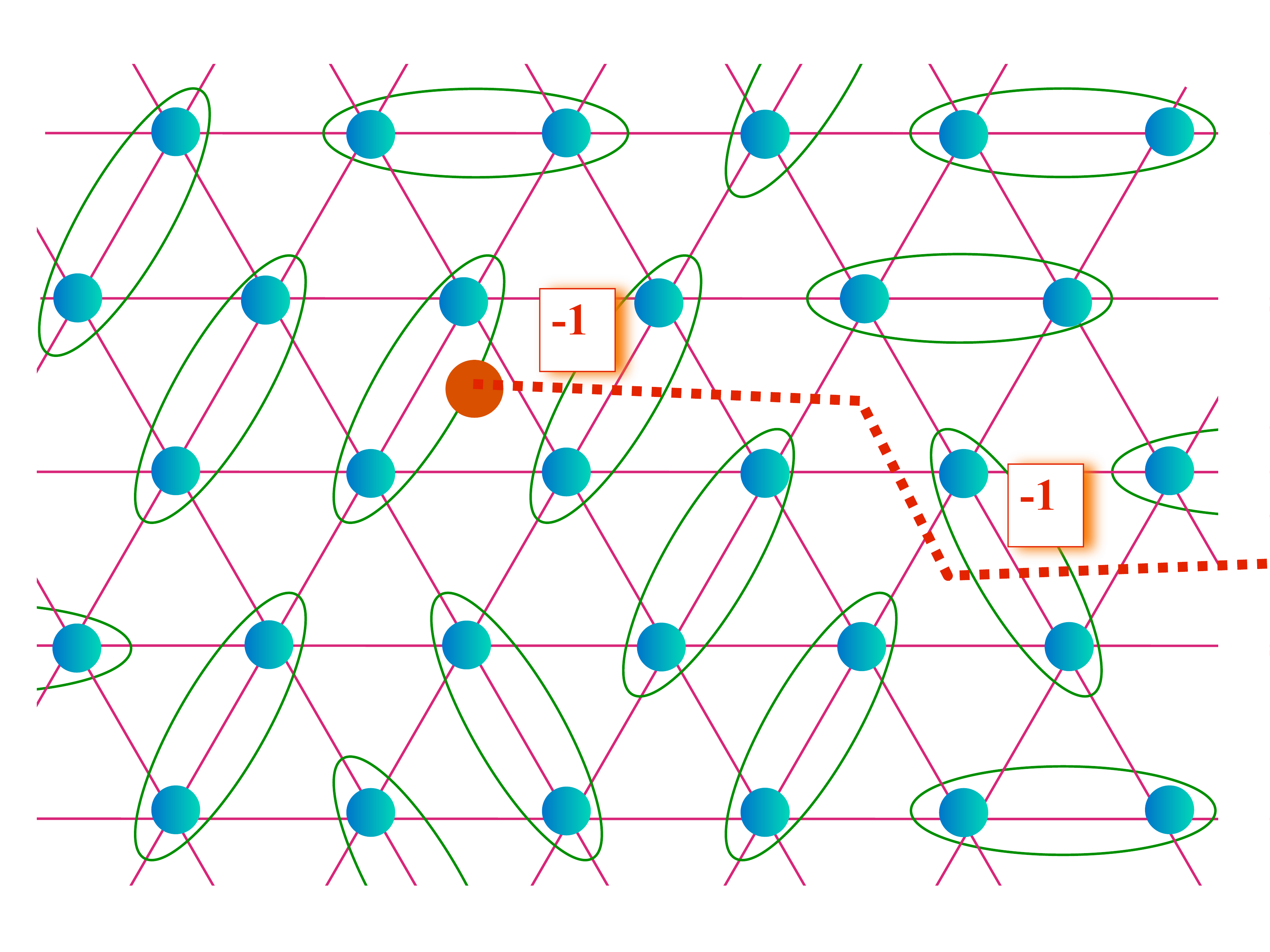}
 \caption{The gapped `vison' excitation of the $\mathbb{Z}_2$ RVB state. This is an excited state whose wavefunction is similar to the
 RVB ground state of Fig.~\ref{fig:afmrvb}. However, the linear superposition over different valence bond configurations has changes in the signs of the terms: the sign is determined by the parity of the number of valence bonds which intersect a `branch-cut' (the dashed red line)
 emanating from the center of the vison. The properties of the vison are independent of the specific location of the branch-cut, which can
 be viewed as a gauge choice. There is $\mathbb{Z}_2$ gauge flux only in the plaquette indicated by the red circle.}
\label{fig:vison}
\end{center}
\end{figure}

Our understanding of the physics of RVB states advanced rapidly after the discovery of cuprate high temperature
superconductivity in 1986. Baskaran and Anderson\cite{bza} 
pointed out that a natural language for the description of RVB-like states is provided
by lattice gauge theory: the constraint on the Hilbert space of one electron per site can be mapped onto the Gauss law constraint of lattice gauge theory. This mapping implies that RVB states can 
also have neutral, spinless excitations which are the analogs of the `photon' of gauge
theories. However, for this picture of the RVB state to hold, it is required that the gauge theory have a stable deconfined phase in which
the spinons can be considered as nearly free particles. Rokhsar and Kivelson\cite{rokhsar} described RVB physics in terms of the
`quantum dimer' model, and discovered a remarkable solvable point at which the simplest RVB state, the equal superposition of 
all nearest-neighbor singlet pairings, was the exact ground state.
Fradkin and Kivelson\cite{fradkiv} showed that the quantum dimer model on a bipartite lattice 
was equivalent to a certain compact U(1) lattice gauge theory. 
However, it remained unclear whether the solvable RVB state was a special
critical point, or part of a RVB phase. It was subsequently argued\cite{rsb} that such U(1) RVB states are generically 
unstable to confinement transitions to states in which the valence bonds crystallize in periodic patterns 
(now called valence bond solids (VBS); see Fig.~\ref{fig:dcp} later).
A stable RVB phase first appeared\footnote{I also note the chiral spin liquid\cite{laughlin,wwz}, 
which breaks time-reversal symmetry spontaneously.
It is closer in spirit to quantum Hall states to be noted later, than to the RVB which breaks no 
symmetries.} in independent works by Wen,\cite{wen1} 
and Read and the author\cite{rstl,sr}, 
who identified it as a deconfined phase of a discrete $\mathbb{Z}_2$ gauge theory\cite{jalabert,sf,sondhi,fendley}. 
The quantum dimer model on the triangular and kagome lattices provides examples of  $\mathbb{Z}_2$ RVB phases,\cite{sstri}
and includes the exactly solvable model as a generic point within the phase.\cite{moessner,pasquier,yao}
The $\mathbb{Z}_2$ RVB state has a gap to all excitations, and so this is 
a realization of gapped quantum matter with long-range entanglement. The analog of the photon in this discrete gauge theory is a gapped
topological excitation known as a `vison', and is illustrated in Fig.~\ref{fig:vison}.
This is a vortex-like excitation, and can propagate across the antiferromagnet like any point particle. It carries neither spin nor charge and 
only energy,
and so is `dark matter'. 

\begin{figure}[htbp]
\begin{center}
 \includegraphics[width=2.5in]{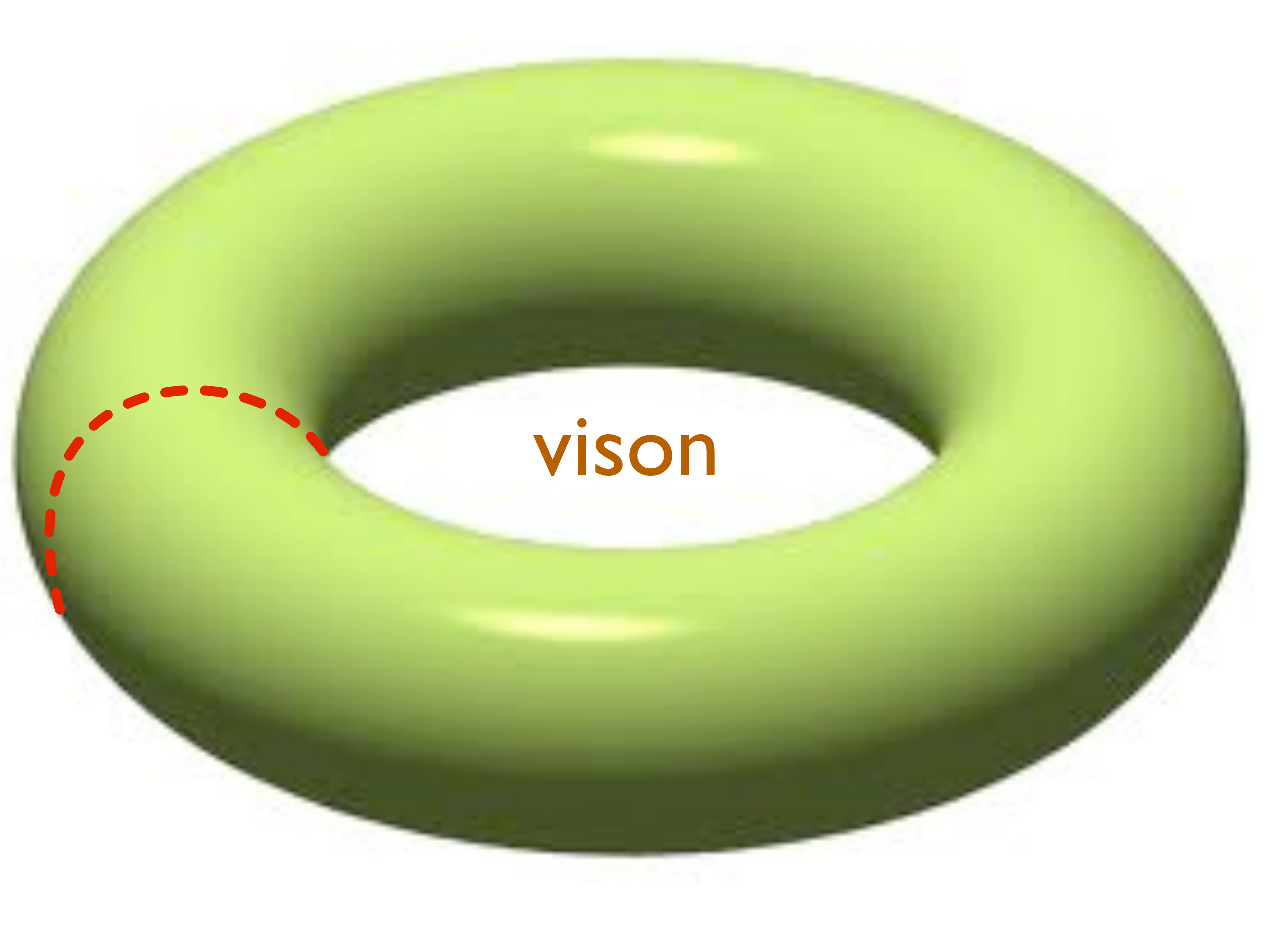}
 \caption{Topological ground state degeneracy of the $\mathbb{Z}_2$ RVB state. The triangular lattice antiferromagnet is placed on the surface of the torus. The dashed line is a branch cut as in Fig.~\ref{fig:vison}.
 The $\mathbb{Z}_2$ gauge flux is now contained in the hole of the torus and so has little influence on the spins.}
\label{fig:torus}
\end{center}
\end{figure}

One of the important consequences of the existence of the vison is that the degeneracy of the RVB state depends upon the topology of the manifold upon which the spins reside: hence it is often stated that the RVB state has `topological order'. Imagine placing the triangular lattice antiferromagnet on the torus, as shown in Fig.~\ref{fig:torus}. 
Then we can arrange the branch-cut of the vison so that the $\mathbb{Z}_2$ gauge flux penetrates one of the holes of the torus.
For a sufficiently large torus, this gauge flux has negligible effect on the energy, and so leads to a two-fold ground state degeneracy.
We can also place the $\mathbb{Z}_2$ in the other hole of the torus, and so the ground state is four-fold 
degenerate.\cite{KRS,rokhsar,thouless,rc}

This ground state degeneracy of the $\mathbb{Z}_2$ RVB state can be viewed as a reflection of its long-range entanglement.
Note that all spin-spin correlation functions decay exponentially fast in the ground state. Nevertheless, there are EPR-type 
long-range correlations by which the quantum state `knows' about the global topology of the manifold on which it resides.
The non-trivial entanglement is also evident in
Kitaev's solvable models\cite{kitaev1,kitaev2} which realize $\mathbb{Z}_2$
spin liquids.

\begin{figure}[htbp]
\begin{center}
 \includegraphics[width=2.5in]{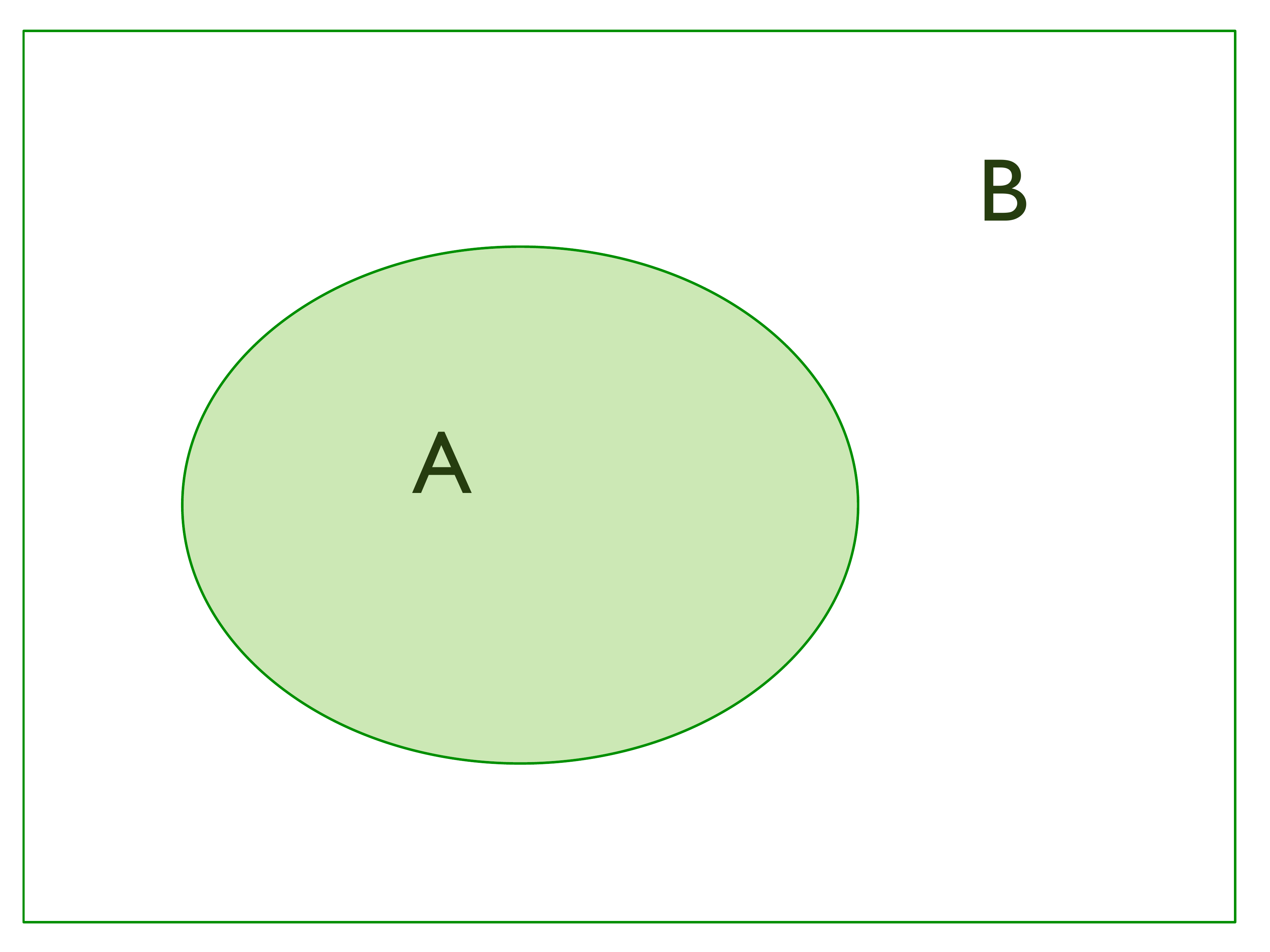}
 \caption{The entanglement entropy of region A is defined by tracing over all the degrees of freedom in region B,
 and computing the von Neumann entropy of the resulting density matrix.}
\label{fig:ee}
\end{center}
\end{figure}

Another measure of the long-range entanglement is provided by the behavior of the entanglement entropy, $S_{E}$. The
definition of this quantity is illustrated in Fig.~\ref{fig:ee}.
We divide the triangular lattice antiferromagnet into two spatial regions, A and B. Then we trace over the spins in region
B and obtain the density matrix $\rho_A = \mbox{Tr}_B \rho$, where $\rho = |\Psi \rangle \langle \Psi |$, with $|\Psi\rangle$
the ground state of the full triangular lattice. The entanglement entropy is $S_E = -\mbox{Tr}\left( \rho_A \ln \rho_A \right)$.
A fundamental feature of the entanglement entropy is that for gapped quantum matter it is expected to obey the `area law';
for the present two-dimensional quantum system, this is the statement\footnote{Eq.~(\ref{see}) is defined in the limit $P\rightarrow \infty$, taken for a fixed shape of the region A; with such a limit, there is no ambiguity in the definition of $\gamma$.}
\beq
S_E = a P - \gamma \label{see}
\eeq
where $P$ is the perimeter of the boundary between regions A and B.  The constant $a$ depends upon microscopic details of the system
under consideration, and is not particularly interesting. Our attention is focused on the value of the offset $\gamma$: this is believed to
provide a universal characterization of the entanglement of the quantum state.  For a band insulator, the entanglement can only depend upon local physics near the boundary,\cite{tarun2} and it is expected that $\gamma =0$. For the $\mathbb{Z}_2$ RVB
state, it was found\cite{preskill,levinwen,tarun1} 
that $\gamma = \ln (2)$: this value of $\gamma$ is then a signature of the long-range entanglement in this state
of gapped quantum matter.

These topological aspects of the $\mathbb{Z}_2$ RVB state can be made more explicit by a mapping of the $\mathbb{Z}_2$ gauge
theory to a doubled Chern-Simons gauge theory.\cite{bais1,bais2,seiberg,freedman,cenke} 
The latter is a topological field theory, and there is a direct connection between
its properties and those of the $\mathbb{Z}_2$ RVB state. Indeed the 4-fold ground state degeneracy on a torus, and the value of the offset $\gamma$ in the entanglement entropy can also be computed in this Chern-Simons theory.\cite{flammia,igor1}

A good candidate for a $\mathbb{Z}_2$ RVB state is the kagome antiferromagnet\cite{sskagome,whitekagome,punk,messio}.
A very recent numerical study \cite{log2} has provided remarkable 
conclusive evidence for the constant $\gamma=\ln (2)$ in the entanglement entropy.
And neutron scattering experiments on such an antiferromagnet display clear signatures of deconfined
spinon excitations.\cite{younglee} There is also compelling evidence for fractionalization and topological order in an
easy-axis kagome antiferromagnet.\cite{bfg,isakov1,isakov2} And finally, several recent studies \cite{j1j20,j1j21,j1j22,sorella}  have argued 
for a gapped spin liquid
in a frustrated square lattice antiferromagnet.

Another large set of  widely studied examples of gapped quantum matter states are the quantum Hall states, and the
related chiral spin liquids.\cite{laughlin,wwz}
We will not discuss these here, apart from noting their relationship to the $\mathbb{Z}_2$ RVB states.
The quantum Hall states do not respect time-reversal invariance and so their topological properties can
be described by Chern-Simons theories with a single gauge field (in the simplest cases). Like the $\mathbb{Z}_2$ RVB states, they 
have ground state degeneracies on a torus, and non-zero values of the entanglement entropy offset $\gamma$.
The quantum Hall states also generically have gapless edge excitations which play a crucial role in their 
physical properties, and such gapless states are not present in the simplest $\mathbb{Z}_2$ RVB state we have discussed above.

\newpage
\section{Conformal quantum matter}
\label{sec:cft}

\begin{figure}[htbp]
\begin{center}
 \includegraphics[width=3.8in]{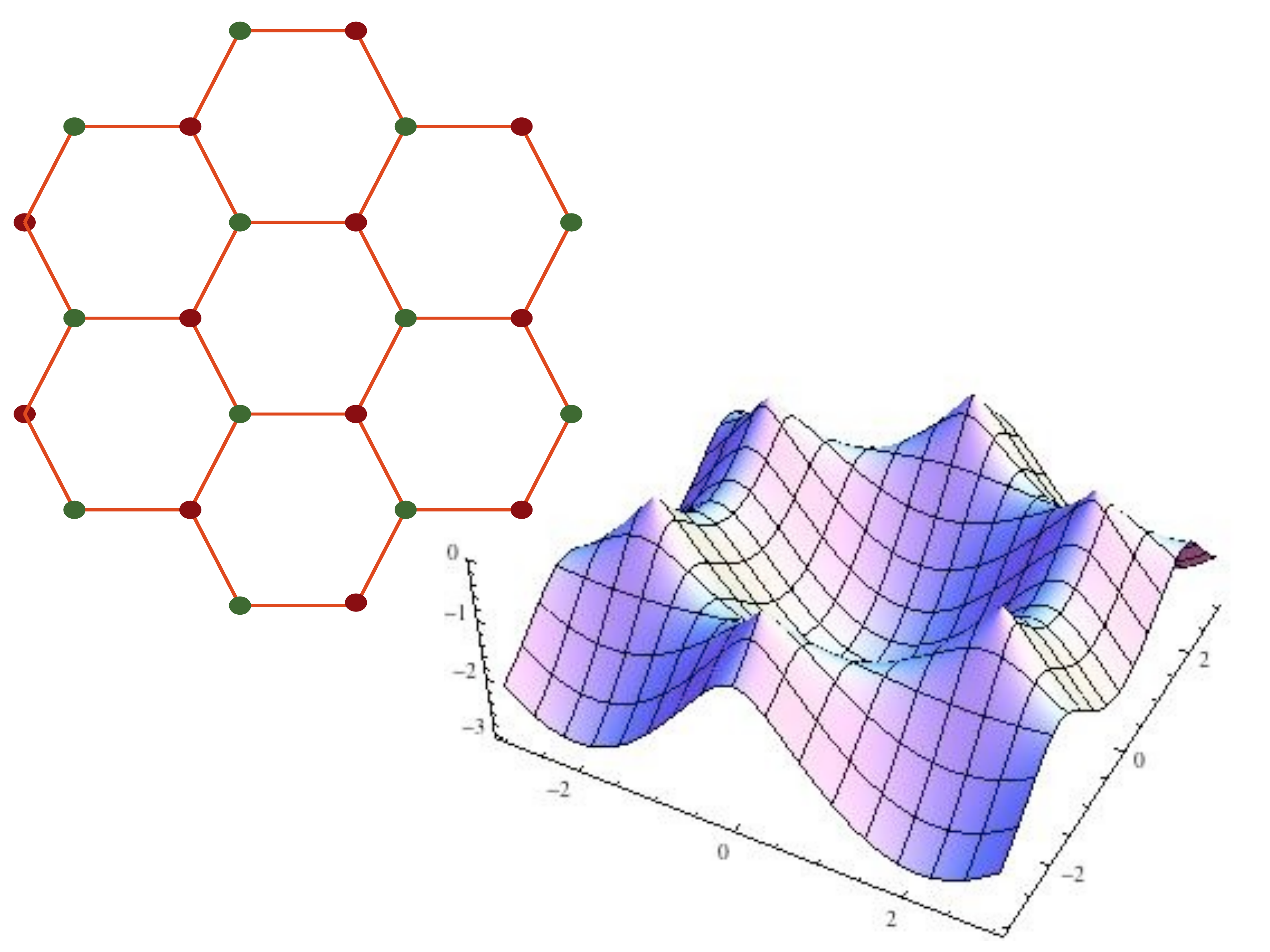}
 \caption{The carbon atoms in graphene (top). The $\pi$ orbitals on the carbon atoms from a half-filled
 band, the lower half of which is shown (bottom). Notice the Dirac cones at six points in the Brillouin zone.
 Only two of these points are inequivalent, and there is a two-fold spin degeneracy, and so 4 two-component
 massless Dirac fermions constitute the low energy spectrum.}
\label{fig:graphene}
\end{center}
\end{figure}

This section considers phases of matter which have gapless excitations at isolated points in the Brillouin zone.
A simple recent example is graphene, illustrated in Fig.~\ref{fig:graphene}.
This has a low energy spectrum of 4 massless Dirac fermions. These fermions interact with the instantaneous Coulomb interaction,
which is marginally irrelevant at low energies, and so the Dirac fermions are free. The theory of Dirac fermions is conformally invariant,
and so we have a simple realization of a conformal field theory in 2+1 spacetime dimensions: a CFT3.
More recently Dirac fermions have also appeared, in both theory and experiment, on the boundary of topological insulators.

\begin{figure}[htbp]
\begin{center}
 \includegraphics[width=4in]{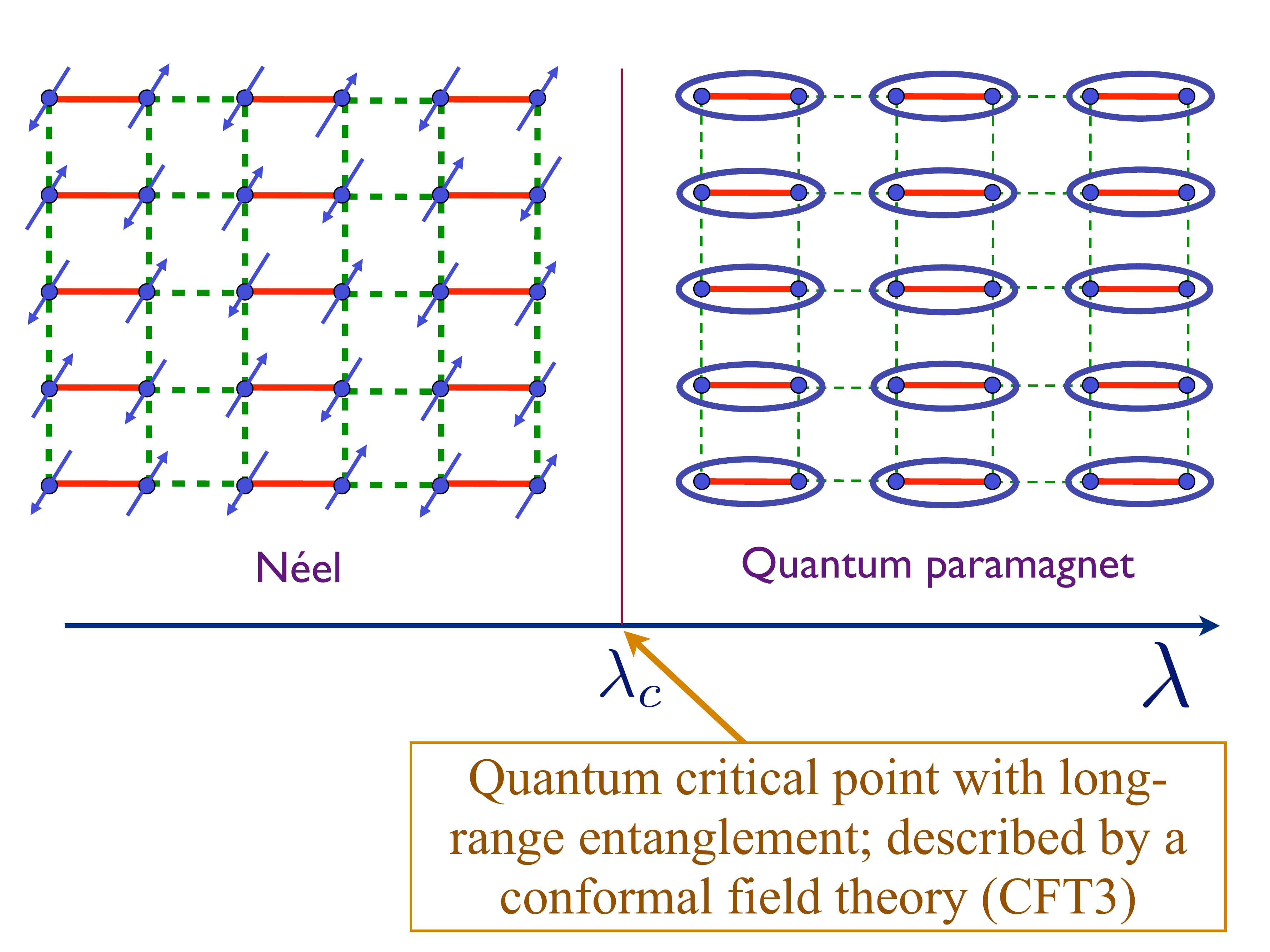}
 \caption{The coupled dimer antiferromagnet. The Hamiltonian is as in Eq.~(\ref{haf}), with the red bonds of strength $J$,
 and the dashed green bonds of strength $J/\lambda$ ($J>0$, $\lambda \geq 1$).}
\label{fig:dimer}
\end{center}
\end{figure}

However, our primary interest is in strongly-interacting CFTs, which provide realizations of quantum matter with long-range
entanglement. One thoroughly studied example is provided by the coupled dimer antiferromagnet, illustrated in Fig.~\ref{fig:dimer}.
This is described by the nearest-neighbor Heisenberg antiferromagnet in Eq.~(\ref{haf}), but with two values of the exchange interactions,
with ratio $\lambda$. For large $\lambda$, the system decouples into dimers, each of which has a spin-singlet valence bond.
This is the quantum paramagnet, which preserves all symmetries of the Hamiltonian and has a gap to all excitations. 
On the other hand, for $\lambda$ close to unity, we obtain a N\'eel state with long-range antiferromagnetic order, similar to that in Fig.~\ref{fig:afmtri}. Both these states have short-range entanglement, and are easily understood by adiabatic continuity from the 
appropriate decoupled limit. However, in between these states is a quantum critical point at $\lambda=\lambda_c$.
There is now compelling numerical evidence\cite{janke} 
that this critical point is described by the CFT3 associated with the Wilson-Fisher fixed point of an interacting field theory of a relativistic scalar with 3 components. Thus a simple generic Heisenberg antiferromagnet flows at low energy to a fixed point
with not only relativistic, but also conformal, invariance.

A notable feature of this CFT3, and of others below, is that it has long-range entanglement in the same sense as that defined
for gapped quantum matter via Eq.~(\ref{see}). The constant $\gamma$ is non-zero \cite{mfs}, and is a characteristic 
universal property of the CFT3 which depends only on the nature of the long-distance geometry, but not its overall scale.

\begin{figure}[htbp]
\begin{center}
 \includegraphics[width=4.4in]{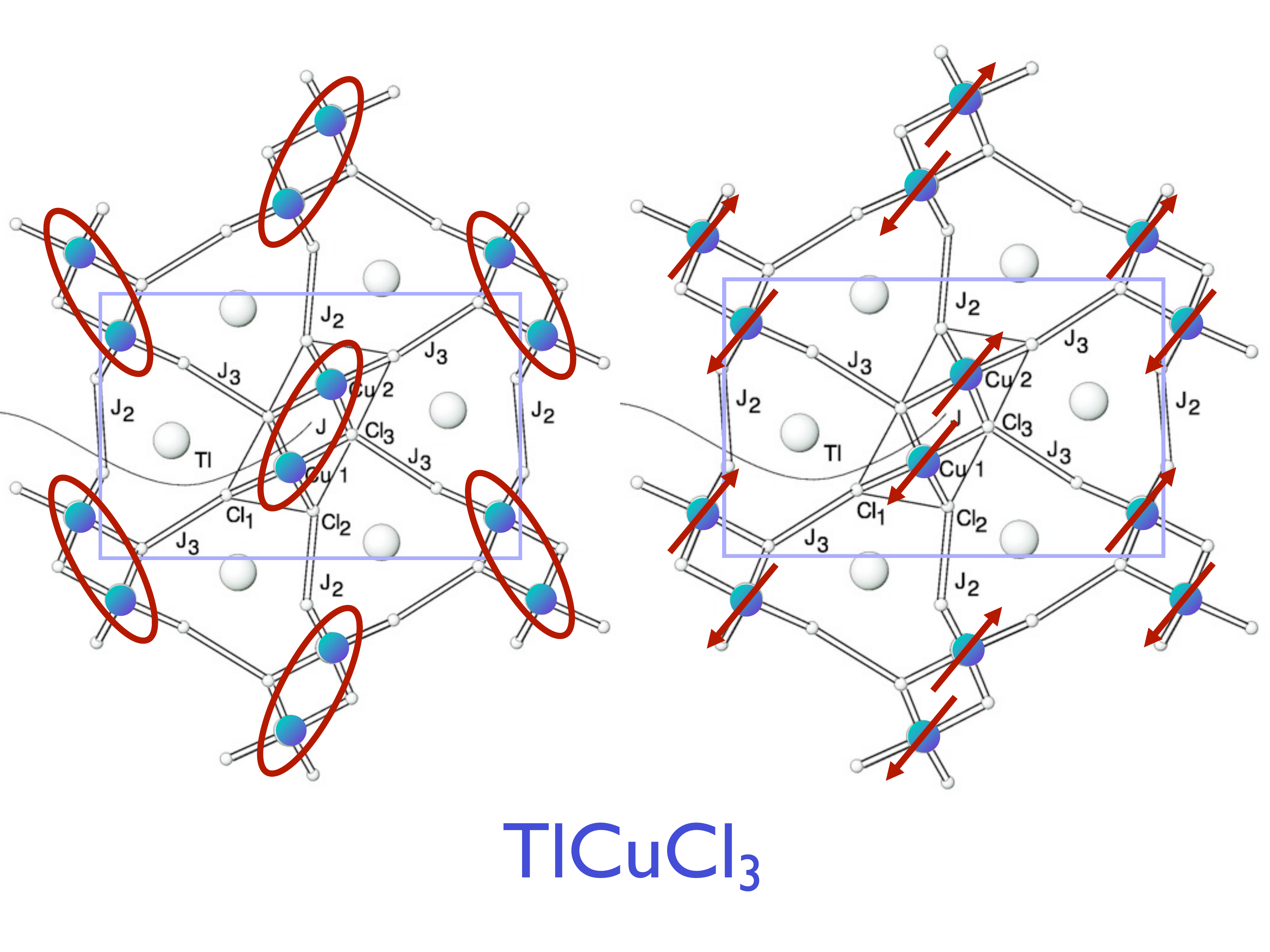}
 \caption{Quantum phase transition in TlCuCl$_3$ induced by applied pressure. Under ambient pressure, TlCuCl$_3$ is a
 gapped quantum paramagnet with nearest-neighbor singlet bonds between the $S=1/2$ spins on the Cu sites (left). Under applied pressure,
 long-range N\'eel order appears (right).}
\label{fig:tlcucl3}
\end{center}
\end{figure}

\begin{figure}[htbp]
\begin{center}
 \includegraphics[width=4in]{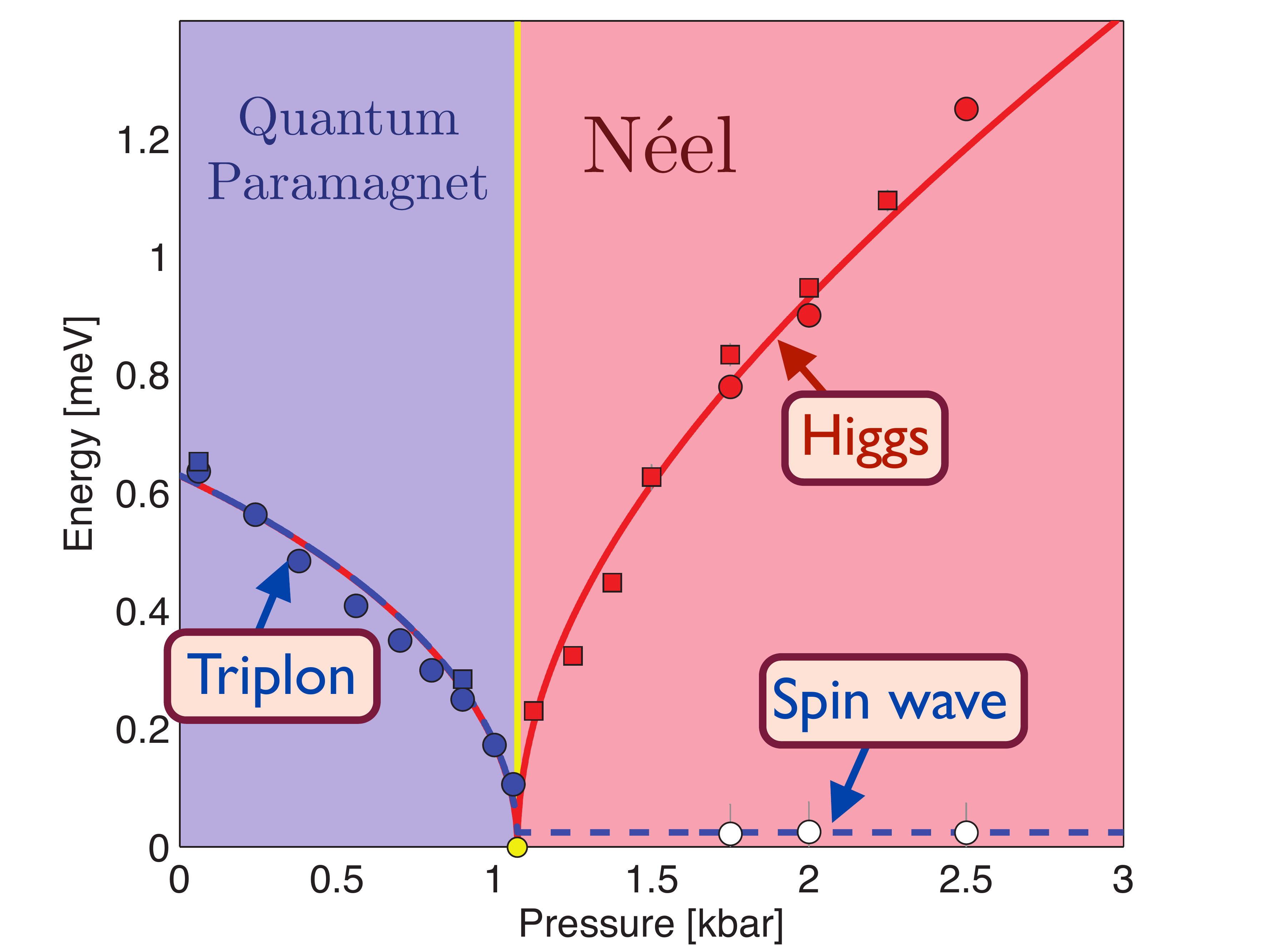}
 \caption{Neutron scattering measurements\cite{ruegg} of the excitation spectrum across the quantum phase transition in TlCuCl$_3$.
 The quantum paramagnet has ``triplon'' excitations, correspond to $S=1$ triplet sets hopping from dimer to dimer.
 The N\'eel state has gapless spin-wave excitations associated with the broken symmetry. It also has a Higgs particle excitation,
 associated with the oscillations in the magnitude of the N\'eel order parameter.}
\label{fig:higgs}
\end{center}
\end{figure}

A possible realization of the coupled dimer antiferromagnet critical point in two dimensions is in Ref.~\refcite{brown}, but detailed
measurements of the excitation spectrum are not available.
However, TlCuCl$_3$ provides nice realization in three dimensions\cite{ruegg}, as shown in Fig.~\ref{fig:tlcucl3}.  
In this case, the quantum-critical point is described by the theory of the 3-component relativistic scalar in 3+1 dimensions; 
the quartic interaction term is marginally irrelevant, and so the critical point is a free CFT4.
The experiments provide an elegant test of the theory of this quantum critical point, as shown in Fig.~\ref{fig:higgs}.
The quantum paramagnet has a `triplon' excitation, which can be interpreted as the oscillation of the scalar field $\vec{\phi}$
about $\vec{\phi} = 0$. The N\'eel phase has gapless spin-wave excitations, which are the Goldstone modes associated with the
broken O(3) symmetry. However, the N\'eel phase also has an excitation\footnote{The Higgs excitation is damped by its ability to emit 
gapless spin waves: this damping is marginal in $d=3$, but much more important in $d=2$.} 
corresponding to oscillations in the magnitude, $|\vec{\phi}|$,
which is the Higgs boson, as discussed in Refs.~\refcite{normand,crossover}. 
Because we are in 3+1 dimensions, we can use mean-field theory to estimate the energies of the
excitations on the two sides of the critical point. A simple mean-field analysis of the potential for $\vec{\phi}$ oscillations
in a Landau-Ginzburg theory shows that\cite{solvay08}
\beq
\frac{ \mbox{Higgs energy at pressure $P = P_c + \delta P$}}{\mbox{Triplon energy at pressure $P=P_c - \delta P$}} = \sqrt{2},
\eeq
where $P_c$ is the critical pressure in Fig.~\ref{fig:higgs}, and $\delta P$ is a small pressure offset. This ratio is well obeyed\cite{solvay08} 
by the data
in Fig.~\ref{fig:higgs}

\begin{figure}[htbp]
\begin{center}
\includegraphics[width=4.5in]{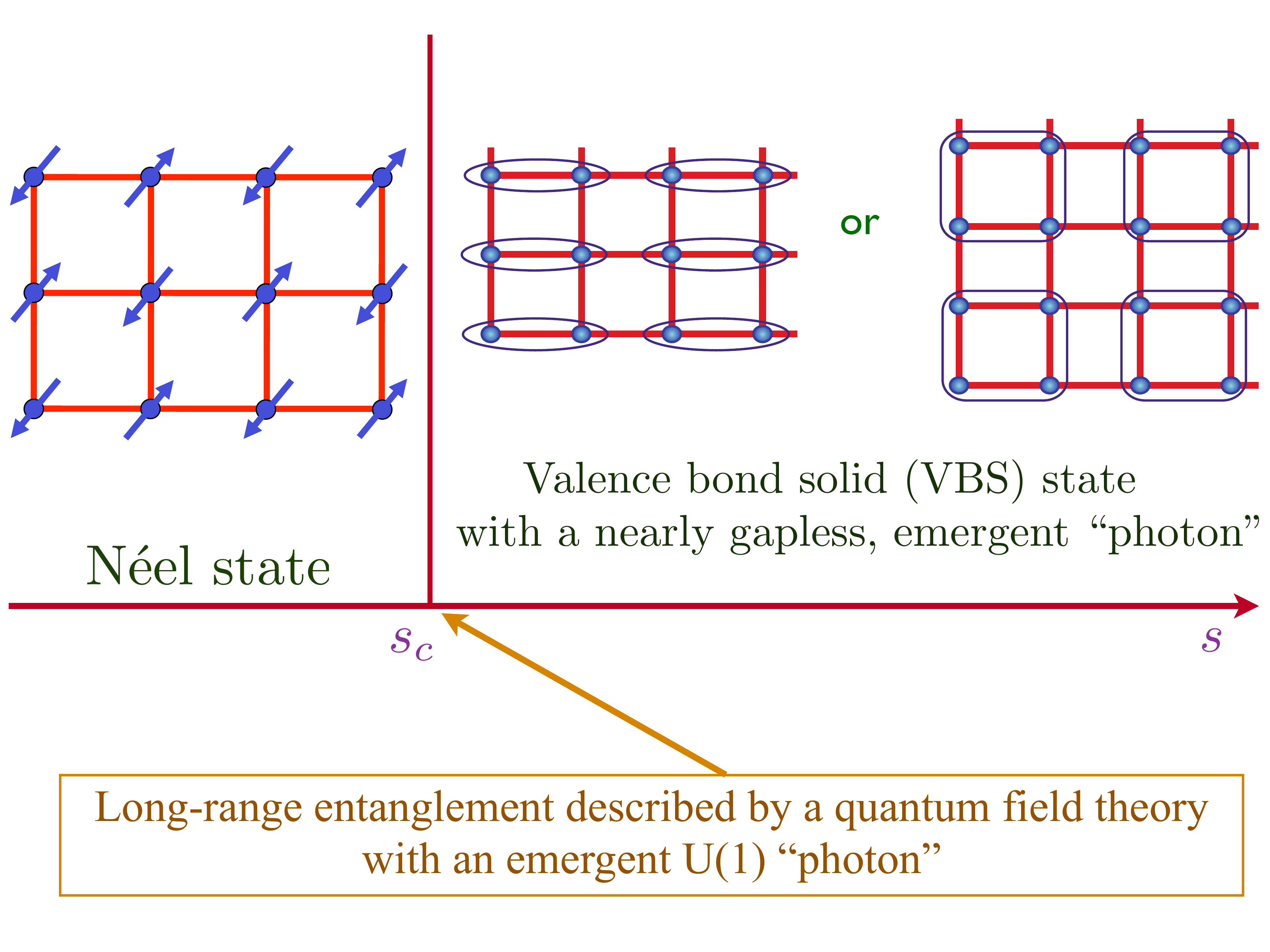}
\caption{Possible phases of a square lattice antiferromagnet, tuned by additional frustrating interactions which are controlled by
the parameter $s$. The N\'eel state breaks spin rotation symmetry. The VBS state breaks lattice symmetry by 
modulating the amplitudes of singlet bonds on the various links of the lattice.}
\label{fig:dcp}
\end{center}
\end{figure}

\begin{figure}[htbp]
\begin{center}
 \includegraphics[width=3in]{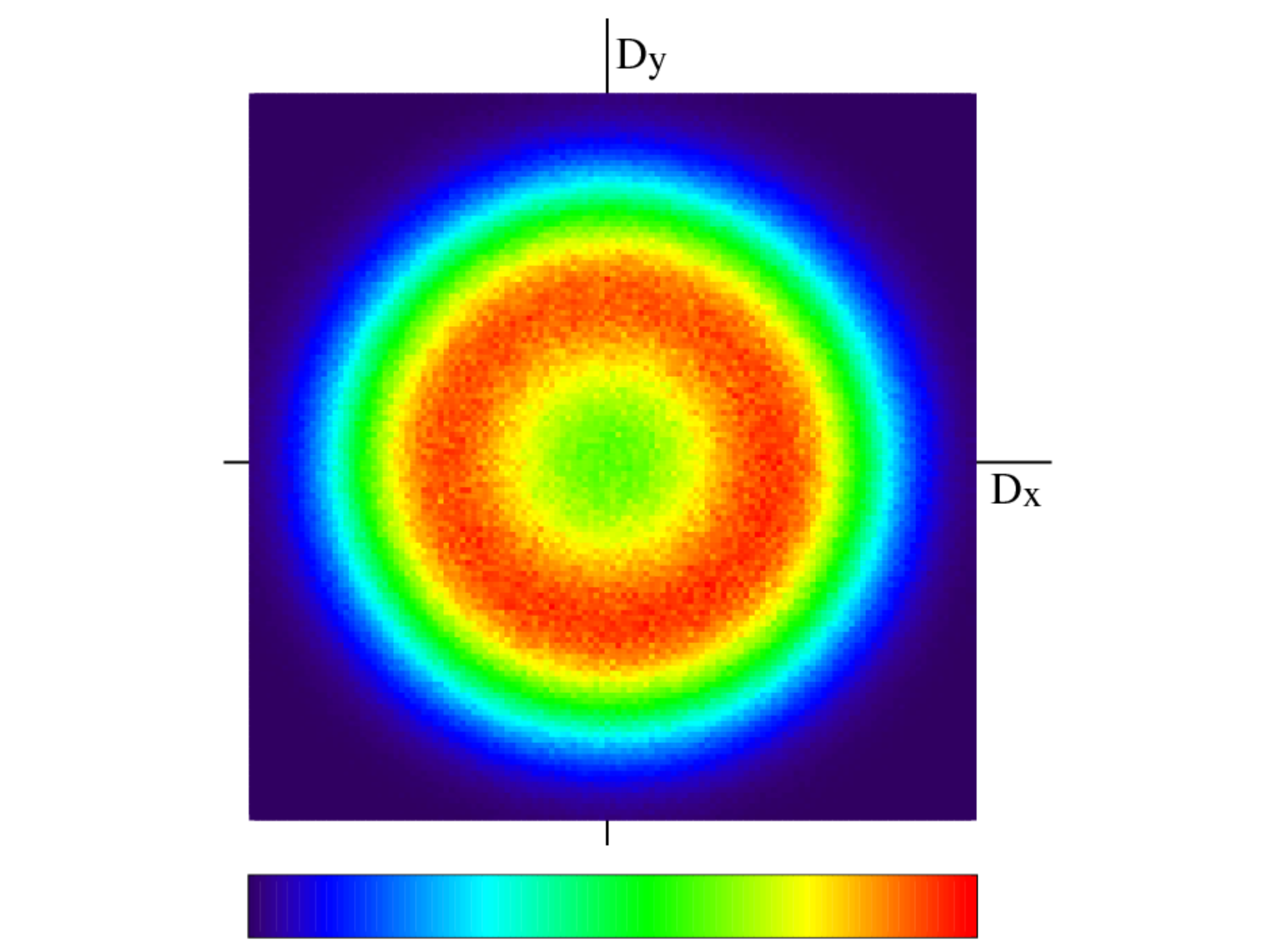}
 \caption{Results from the studies of a square lattice antiferromagnet by Sandvik.\cite{sandvik} The measurements are at the $s=s_c$
 critical point between the N\'eel and VBS states of Fig.~\ref{fig:dcp}. $D_x$ is a measure of the VBS order along the $x$ direction:
 $D_x = \sum_{j} (-)^{j_x} \vec{S}_j \cdot \vec{S}_{j+ \hat{e}_x}$, and similarly for $D_y$; here $j \equiv (j_x, j_y)$ labels square lattice sites, and $\hat{e}_x$ is a unit vector in the $x$ direction. The emergent circular symmetry of the distribution of $D_x$ and $D_y$
 is evidence for the existence of a gapless scalar field, which is the dual of the emergent U(1) photon.\cite{senthil2,sandvik}}
\label{fig:photon}
\end{center}
\end{figure}

The quantum antiferromagnet of Fig.~\ref{fig:dimer} is special in that it has two $S=1/2$ per unit cell: this makes the structure
of the quantum paramagnet especially simple, and allows for description of the quantum critical point by focusing on the fluctuations
of the N\'eel order alone. The situation becomes far more complex, with new types of CFT3s, 
when we consider models with a single $S=1/2$ per unit cell. A prominent example is the
frustrated square lattice antiferromagnet. With only nearest neighbor interactions, the square lattice antiferromagnet has long-range N\'eel order, as illustrated on the left side of Fig.~\ref{fig:dcp}.
After applying additional interactions which destablize the N\'eel state, but preserve full square lattice symmetry, certain antiferromagnets
exhibit a
quantum phase transition to a VBS state which restores spin rotation invariance but breaks lattice symmetries.\cite{rsl}
It has been argued\cite{senthil1,senthil2} 
that this quantum phase transition is described by a field theory of a non-compact U(1) gauge field
coupled to a complex bosonic spinor {\em i.e.\/} a relativistic boson which carries unit charge of the U(1) gauge field and transforms as a 
$S=1/2$ fundamental of the global SU(2) spin symmetry (note to particle theorists: here ``spin'' refers to a global symmetry analogous to 
flavor symmetry, and so there are no issues with the spin-statistics theorem). Evidence for this proposal has appeared in numerical studies by Sandvik\cite{sandvik}, as illustrated in Fig.~\ref{fig:photon}, which shows remarkable evidence for an `emergent photon'.
It is possible that the experimental system of Ref.~\refcite{brown} exhibits a N\'eel-VBS transition.

A separate question is whether the critical point of this theory of a non-compact U(1) photon coupled to the relativistic boson
is described by a CFT3. The existence of such a `deconfined critical point' has been established by a $1/N$ expansion, in a model in which the global SU(2) symmetry is enlarged to SU($N$). Recent numerical studies\cite{ribhu} also show strong support for the existence of the deconfined critical theory for $N>4$. The important $N=2$ case has not been settled, although there is now 
evidence for a continuous transition
with rather slow transients away from the scaling behavior.\cite{sandvik2,damle2}

\begin{figure}[htbp]
\begin{center}
 \includegraphics[width=3.6in]{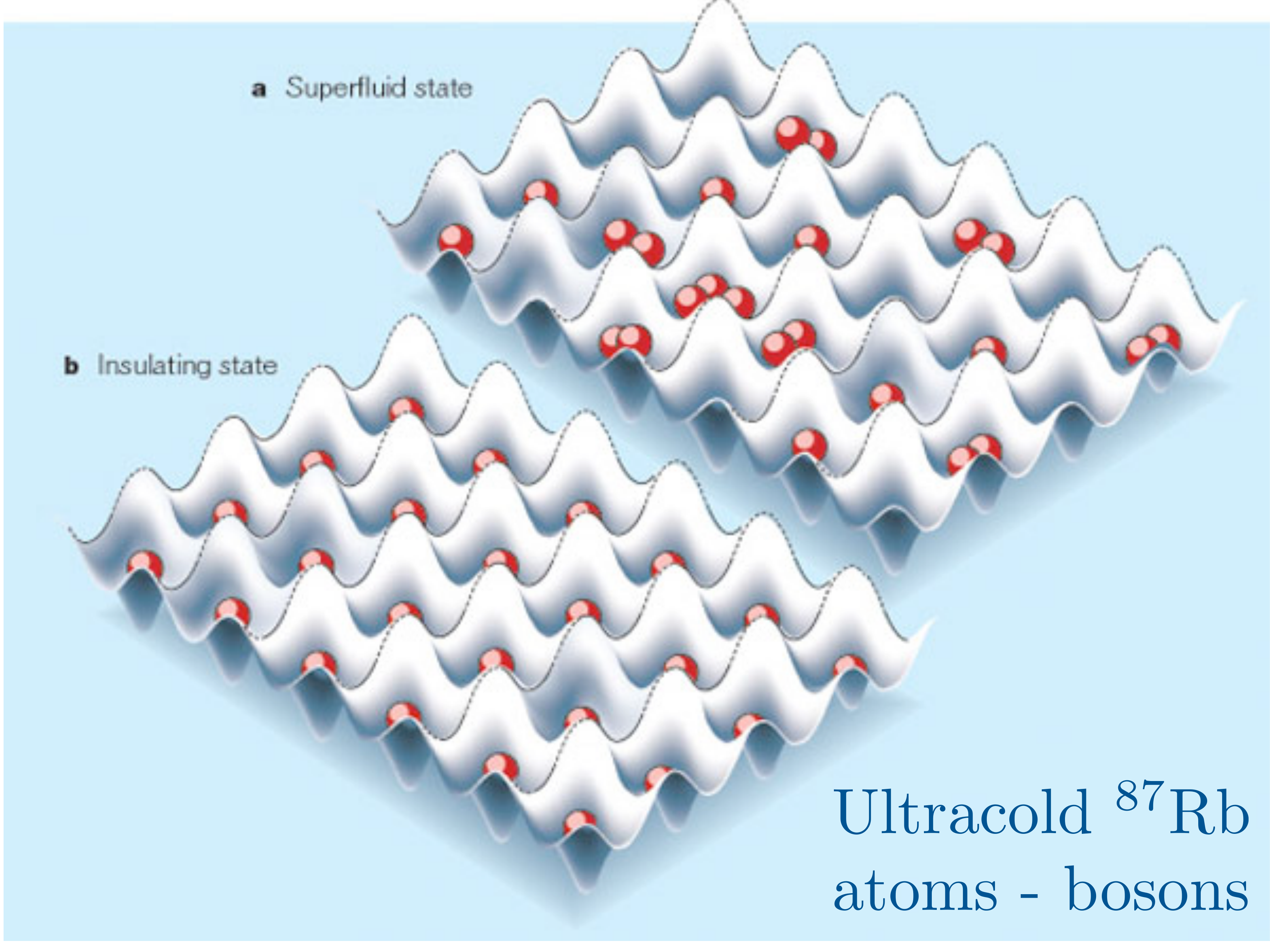}
 \caption{Bosons in an optical lattice undergo a superfluid-insulator transition as the depth of the optical lattice
 is increased, when there is an integer density of bosons per site. The critical theory is described by a relativistic field
 theory of a complex scalar with short-range self interactions.}
\label{fig:hubbard}
\end{center}
\end{figure}

In the near future, ultracold atoms appear to be a promising arena for experimental studies of CFT3s.
Bosonic $^{87}$Rb atoms were observed to undergo such a quantum phase transition in an optical lattice\cite{bloch}, 
as shown in Fig.~\ref{fig:hubbard}.
The quantum critical point here is described by the same relativistic scalar field theory as that discussed above for the dimer antiferromagnet,
but with the field $\vec{\phi}$ now having two components with a global O(2) symmetry linked to the conservation of boson number.\cite{fwgf}
This critical point is described by the Wilson-Fisher fixed point in 2+1 dimensions, which realizes a strongly interacting CFT3.
Experiments on the superfluid-insulator transition in two dimensions have now been performed,\cite{chin} and this opens the way towards a detailed
study of the properties of this CFT3. In particular, the single-site resolution available in the latest experiments\cite{simon,kuhr} 
promises detailed information on real time dynamics with detailed spatial information.

\begin{figure}[htbp]
\begin{center}
 \includegraphics[width=4in]{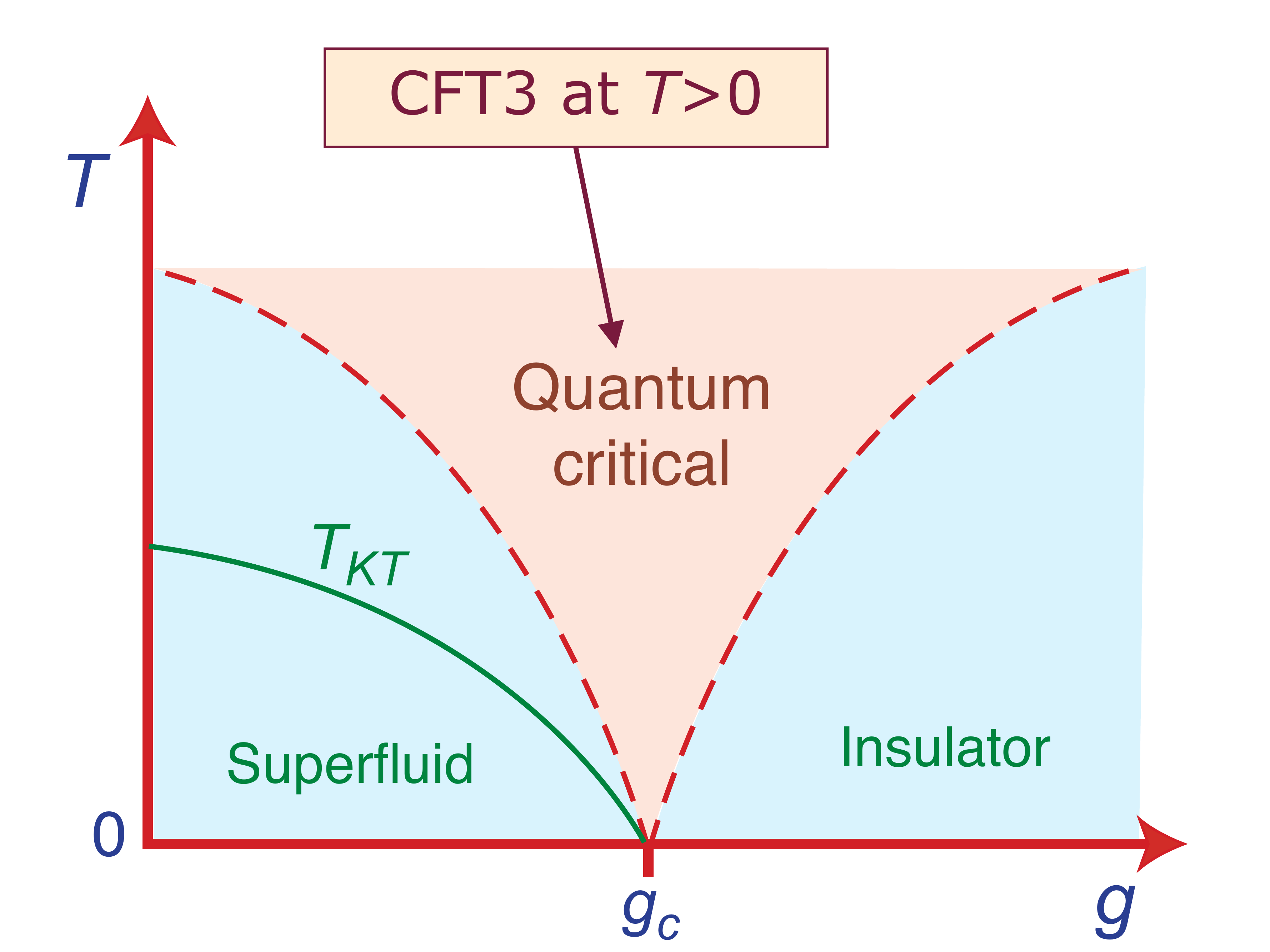}
 \caption{Non-zero temperature phase diagram of the superfluid-insulator transition in two spatial dimensions.
 Quasi-long-range superfluid order is present below the Kosterlitz-Thouless transition temperature $T_{KT}$. 
 The dashed lines are locations of crossovers.}
\label{fig:qc}
\end{center}
\end{figure}

These and other experiments demand an understanding of the real time dynamics of CFT3s at non-zero temperatures ($T$).
We sketch the nature of the $T>0$ phase diagram for the superfluid-insulator transition in Fig.~\ref{fig:qc}.
In the blue regions, the long time dynamics are amenable to a classical description:
in the ``superfluid'' region we can use the Gross-Pitaevski non-linear wave equation, while in the ``insulator'' region we
can describe the particle and hole excitations of the insulator using a Boltzmann equation. However, most novel is
the pink ``Quantum Critical'' region,\cite{chn} where classical models cannot apply at the longest characteristic time scales.
In fact, in this region, all the characteristic time scales are set by temperature alone, and we have\cite{jinwu1,jinwu2}
\beq
\tau = \mathcal{C} \frac{\hbar}{k_B T}
\eeq
where $\tau$ is some appropriately defined relaxation time, and $\mathcal{C}$ is a universal constant characteristic of the CFT3. 
The computation of $\mathcal{C}$, and related dissipative and transport co-efficients is a challenging task, and is 
not easily accomplished by the traditional expansion and renormalization group methods of quantum field theory. It is in these questions
that the methods of gauge-gravity duality have had some impact, as the author has reviewed elsewhere.\cite{arcmp}

\newpage
\section{Compressible quantum matter}
\label{sec:nfl}

As the name implies, compressible states are those whose ``density'' can be varied freely by tuning an external parameter. 
Remarkably, there are only a few known examples of states which are compressible at $T=0$. On the other hand,
compressible quantum phases are ubiquitous in intermetallic compounds studied in recent years, 
and many of their observable properties 
do not fit into the standard paradigms. 
So a classification and deeper understanding of the possible compressible phases of quantum matter is of considerable
importance.

Let us begin our discussion with a definition of compressible quantum matter.\cite{liza,arcmp}
\begin{itemize}
\item Consider a continuum, translationally-invariant 
quantum system with a globally conserved U(1) charge $\mathcal{Q}$
{\em i.e.\/} $\mathcal{Q}$ commutes with the Hamiltonian $H$. Couple the Hamiltonian
to a chemical potential, $\mu$, which is conjugate to $\mathcal{Q}$: so the Hamiltonian
changes to $H - \mu \mathcal{Q}$. The ground state of this modified Hamiltonian
is compressible if $\langle \mathcal{Q} \rangle$ changes smoothly as a function of $\mu$, 
with $d\langle \mathcal{Q} \rangle/d\mu$ non-zero.
\end{itemize}
A similar definition applies to lattice models, but let us restrict our attention to continuum models for simplicity.

Among states which preserve both the translational and global U(1) symmetries, the only traditional condensed matter state
which is compressible is the Fermi liquid. This is the state obtained by turning on interactions adiabatically on the Sommerfeld-Bloch
state of non-interacting fermions. Note that in our definition of compressible states we have allowed the degrees of freedom
to be bosonic or fermionic, but there are no compressible states of bosons which preserve the U(1) symmetry.

One reason for the sparsity of compressible states is that they have to be gapless. Because $\mathcal{Q}$ commutes with $H$,
changing $\mu$ will change the ground state only if there are low-lying levels which cross the ground state upon an infinitesimal
change in $\mu$. For the gapless states of conformal quantum matter considered in Section~\ref{sec:cft}, a scaling argument
implies a compressibility $\sim T^{d-1}$. So such states are compressible only in $d=1$. The known $d=1$ compressible
states are `Luttinger liquids' or their variants: they have a decoupled massless relativistic scalar with central charge $c=1$ representing
the fluctuations of $\mathcal{Q}$. We will not be interested in such states here.

The key characteristic of the Fermi liquid is the Fermi surface. For interacting electrons, the Fermi surface is defined by a zero
of the inverse fermion Green's function 
\beq
G_f^{-1} ( |{\bm k}| = k_F, \omega =0 ) = 0.  \label{fszero}
\eeq
The Green's function is a complex number, and so naively the variation of the single real parameter $|{\bm k}|$ in Eq.~(\ref{fszero}) does not
guarantee that a solution for $k_F$ is possible. However, we can find $k_F$ by solving for the real part of Eq.~(\ref{fszero}). In all known cases, we find that the imaginary part of $G_f^{-1}$ also vanishes at this $k_F$: this happens because $k_F$ is the momentum where the energy of both particle-like and hole-like excitations vanish, and so there are no lower energy excitations for them to decay to.\footnote{This argument also shows why the equation for the zeros of the Green's function $G_f( |{\bm k}| , \omega=0) = 0$ generically 
has no solution (although they are important for the approach reviewed in Ref.~\refcite{yrz}). 
In this case the vanishing of the real part has no physical interpretation, and the imaginary part need not vanish at the same $|{\bm k}|$.} 

\begin{figure}[htbp]
\begin{center}
 \includegraphics[width=3.3in]{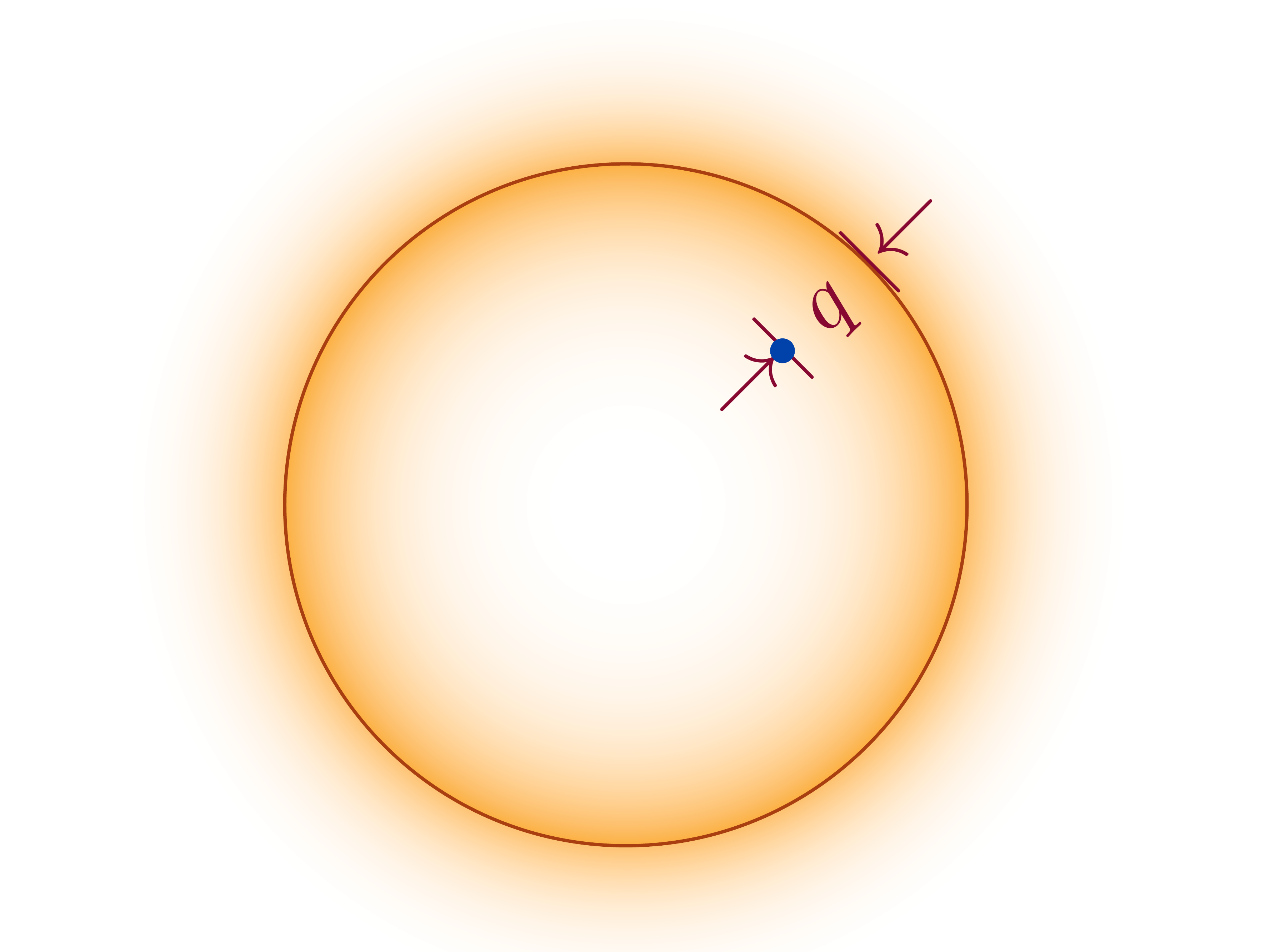}
 \caption{The Fermi surface. The fermion at the blue point is momentum $-q$ away from the nearest point on the Fermi surface.
 There is a sharp quasiparticle pole on the Fermi surface for a Fermi liquid, as in Eq.~(\ref{fl}).
 There are no quasiparticles in non-Fermi liquids, but a continuum of low energy excitations which obey Eq.~(\ref{scale});
 nevertheless, the position of the Fermi surface is well-defined, and it encloses volume which obeys a Luttinger relation.
 For the model in Eq.~(\ref{nfl}), this Fermi surface is {\em hidden}, because the single fermion Green's function in Eq.~(\ref{scale}) is
 not a gauge-invariant observable.}
\label{fig:fs}
\end{center}
\end{figure}

In a Fermi liquid, the Green's function
has a simple pole at the Fermi surface with
\beq
G_f^{-1} = \omega - v_F q + \mathcal{O} (\omega^2, q^2) \label{fl}
\eeq
where $q = |{\bm k}| - k_F$ is the minimal distance to the Fermi surface (see Fig.~\ref{fig:fs}), and $v_F$ is the Fermi velocity.
The relationship between $k_F$ and the density $\langle \mathcal{Q} \rangle$ in a Fermi liquid is the same as
that in the free fermion state: this is the Luttinger relation, which equates $\langle \mathcal{Q} \rangle$ to the 
momentum-space volume enclosed by the Fermi surface (modulo phase space factors).

Numerous modern materials display metallic, compressible states which are evidently not  Fermi liquids.
Most commonly, they are associated with metals near the onset of antiferromagnetic long-range order; these materials
invariably become superconducting upon cooling in the absence of an applied magnetic field.
The onset of antiferromagnetism in metals, and the proximate presence of ``high-$T_c$'' superconductivity,
was discussed by the author at the Solvay conference;
however, the subject has been reviewed in a separate 
recent article\cite{sces}, and so will not be presented here.

\begin{figure}[htbp]
\begin{center}
 \includegraphics[width=4.8in]{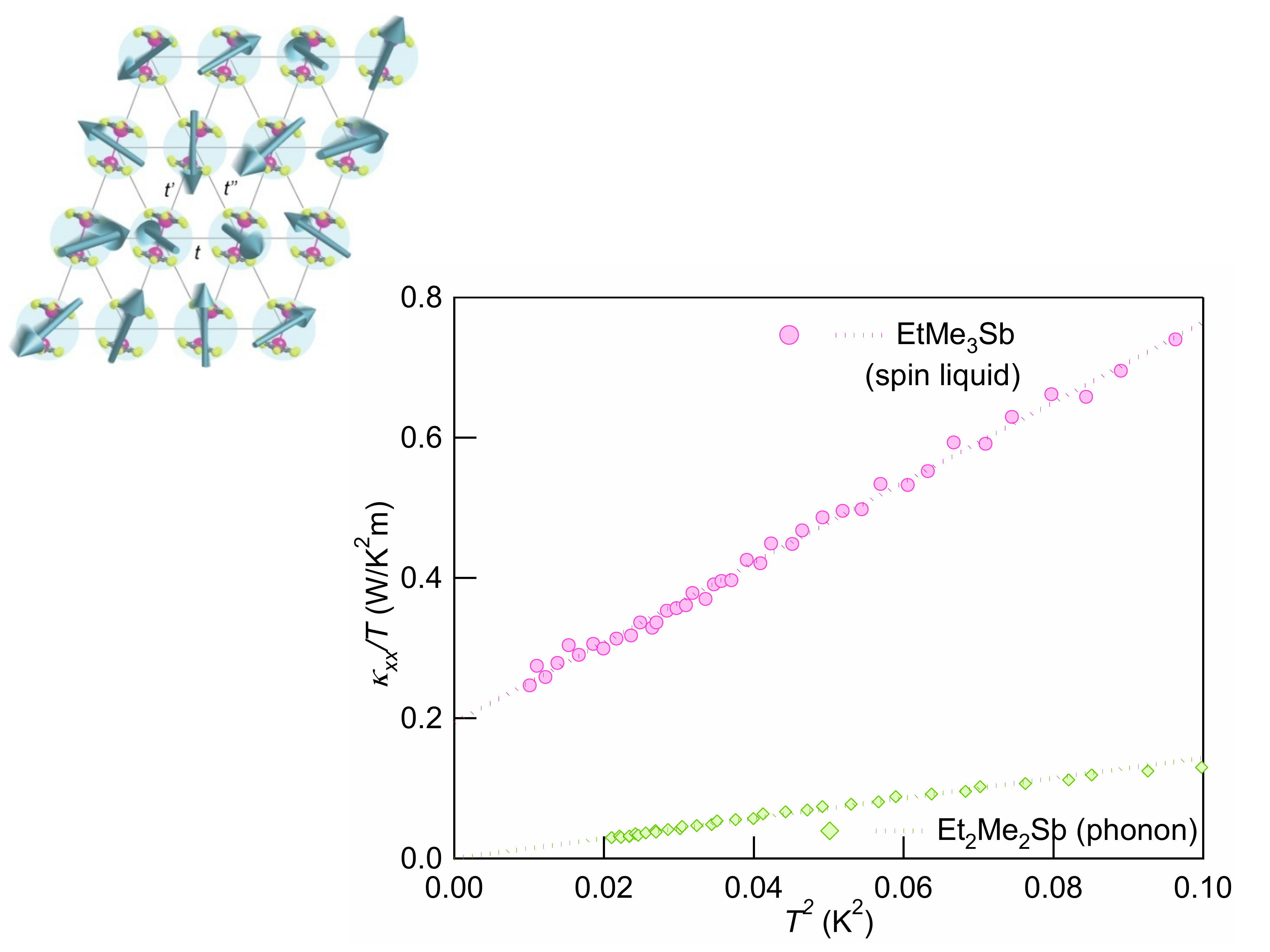}
 \caption{From Ref.~\refcite{etsb}. The longitudinal thermal conductivity $\kappa_{xx}$ as a function
 of temperature ($T$) for EtMe$_3$Sb[Pd(dmit)$_2$]$_2$, an insulating antiferromagnet of $S=1/2$ spins on a triangular 
 lattice (sketched at the top). 
 The notable feature is the non-zero value of $\lim_{T \rightarrow 0} \kappa_{xx}/T$, which is characteristic of thermal transport of fermions near a Fermi surface.}
\label{fig:etsb}
\end{center}
\end{figure}

Here, we note another remarkable compressible phase found in the organic insulator 
EtMe$_3$Sb[Pd(dmit)$_2$]$_2$. This has a triangular lattice of $S=1/2$ spins, as in the antiferromagnet discussed
in the beginning of Section~\ref{sec:gap}; however, there are expected to be further neighbor ring-exchange interactions beyond
the nearest-neighbor term in Eq.~(\ref{haf}), and possibly for this reason the ground state does not have antiferromagnetic order,
and nor does it appear to be the gapped $\mathbb{Z}_2$ RVB state. Remarkably,
the low temperature thermal conductivity of this material
is similar to that of a metal\cite{etsb}, even though the charge transport is that of an insulator: see Fig.~\ref{fig:etsb}.
Thus this material is a charge insulator, but a thermal metal. One possible explanation is that there is a Fermi surface of 
spinons\cite{mot,mot2,mot3,biswas,lai,hermelemajorana}, which
would also be consistent with the observed non-zero spin susceptibility. This Fermi sea of spinons realizes a phase of compressible quantum matter, where the conserved charge $\mathcal{Q}$ is identified with the total spin.

Motivated by these and other experiments, we now turn to a discussion of a much-studied realization of compressible quantum
matter which is not a Fermi liquid. This is the problem of fermions, $\psi$, at non-zero density coupled to an Abelian or non-Abelian gauge field, $A^a$, of a Lie group.\footnote{It is not clear whether such models apply to EtMe$_3$Sb[Pd(dmit)$_2$]$_2$. The theories of Refs.~\refcite{mot,mot2,mot3} 
have continuous
gauge groups, while those of Ref.~\refcite{biswas,lai,hermelemajorana} have 
discrete gauge groups; Eq.~(\ref{nfl}) does not apply to the latter.}
We can schematically write the Lagrangian as
\beq
\mathcal{L} = \psi^\dagger \left( \partial_\tau - i A^a_\tau t^a - \mu h \right) \psi - \frac{1}{2m} 
\psi^\dagger \left( {\bm \nabla} - i {\bm A}^a t^a \right)^2 \psi + \frac{1}{4g^2} F^2 \label{nfl}
\eeq
where $F$ is the field tensor, $\tau$ is imaginary time, $\mu$ is the chemical potential, $t^a$ are the generators of the gauge group, $h$ is the generator of the conserved charge $\mathcal{Q}$
($h$ is distinct from, and commutes with, all the $t^a$), and $m$ is the effective mass. In the application to spin liquids, $\psi$ represents the fermionic spinons, and $A^a$ is the emergent gauge field of a particular RVB state.

Let us summarize the present understanding of the properties of (\ref{nfl}) in spatial dimension $d=2$, obtained
by conventional field-theoretic 
analysis.\cite{reizer,palee,monien,hlr,polchinski,nayak,bim,furusaki,sslee,metnem,mross,metzner,bartosch} 
There is a universal, compressible `non-Fermi liquid' state with a Fermi surface 
at precisely the same $k_F$ as that given by the free electron value. However, unlike the Fermi liquid, this Fermi surface is \underline{\em hidden},
and characterized by singular, non-quasiparticle low-energy excitations. It is hidden because the  
$\psi$ fermion Green's function is not a gauge-invariant quantity, and so is not a physical observable. However, in perturbative theoretic analyses, the $\psi$ Green's function can be computed in a fixed gauge, and this quantity is an important ingredient which determines the singularities of physical observables. In the Coulomb gauge, ${\bm \nabla} \cdot {\bm A}^a = 0$, 
the $\psi$ Green's function has been argued to obey the scaling form\cite{metnem}
\beq
G_\psi^{-1} = q^{1-\eta} \Phi (\omega/q^z) \label{scale}
\eeq
where $q$ is the momentum space distance 
from the Fermi surface, as indicated in Fig.~\ref{fig:fs}. The function $\Phi$ is a scaling function which characterizes
the continuum of excitations near the Fermi surface, $\eta$ is an anomalous dimension, and $z$ is a dynamic critical exponent.
The Fermi liquid result clearly corresponds to $\eta=0$ and $z=1$, and simple form for $\Phi$. For the present non-Fermi liquid, 
the exponent $\eta$ was recently estimated in loop expansions.\cite{metnem,mross} It was also found that $z=3/2$ to 
three loops,\cite{metnem} and it is not known
if this is an exact result.\footnote{In the condensed matter literature, it is often stated that this theory has $z=3$. This refers to the dynamic scaling of the gauge field propagator, which has\cite{metnem} exactly twice the value of $z$ from that defined by Eq.~\ref{scale}}

For our discussion below, we need the thermal entropy density, $S$, of this non-Fermi liquid compressible state at low
temperatures.
This is found to be
\beq
S \sim T^{1/z}. \label{therm}
\eeq
This can be viewed as an analog of the Stefan-Boltzmann law, which states that $S \sim T^{d/z}$ for a $d$-dimensional
quantum system with excitations which disperse as $\omega \sim |{\bm k}|^z$. In the present case, our critical fermion excitations
disperse only transverse to the Fermi surface, and so they have the phase space, and corresponding entropy, of
effective dimension $d_{\rm eff} = 1$. Following critical phenomena terminology, let us rewrite Eq.~(\ref{therm}) in the form
\beq
S \sim T^{(d-\theta)/z}, \label{therm2}
\eeq
where $\theta $ is the violation of hyperscaling exponent, defined by $d_{\rm eff} = d-\theta$. 
The present non-Fermi liquid therefore has
\beq
\theta = d-1. \label{theta}
\eeq

\begin{figure}[htbp]
\begin{center}
 \includegraphics[width=3.3in]{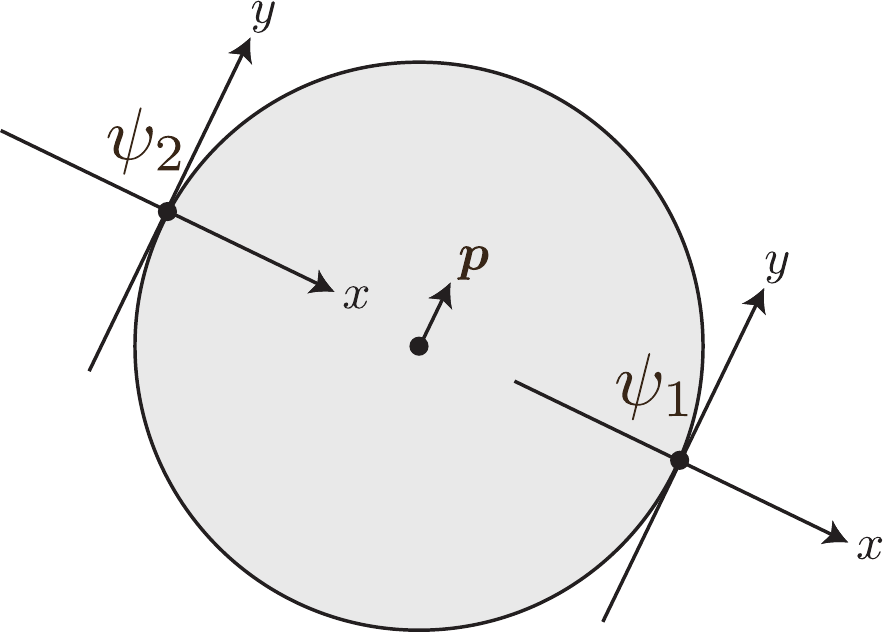}
 \caption{Low energy limit of Eq.~(\ref{nfl}). The $A$ gauge-field fluctuations with momenta ${\bm p}$ couple most
 efficiently to fermions $\psi_{1,2}$ near Fermi surface points where the tangent is collinear to ${\bm p}$.}
\label{fig:tangent}
\end{center}
\end{figure}

We conclude this section by giving a few more details of the derivation of the above scaling properties of Eq.~(\ref{nfl}) for the case
of a U(1) gauge group, focusing on the determination of the value of $z$. In the low energy limit, it has been argued \cite{sslee,metnem}
that we can focus on the gauge field 
fluctuations collinear to single direction ${\bm p}$, and these couple most efficiently to fermions
at antipodal points on the Fermi surface where the tangent to the Fermi surface is also parallel to ${\bm p}$: see Fig.~\ref{fig:tangent}.
From Eq.~(\ref{nfl}), it is then straightforward to derive the following low energy action for the 
long-wavelength fermions, $\psi_{1,2}$ at the antipodal points, and the gauge field $A$
\bea
\mathcal{S} &=& \int d\tau dx dy  \Biggl[ \psi_1^\dagger \left(  \partial_\tau - i \partial_x - \partial_y^2 \right) \psi_1 + \psi_2^\dagger \left(  \partial_\tau + i \partial_x - \partial_y^2 \right) \psi_2 \nn
&~& \quad \quad \quad - g \, A \left( \psi_1^\dagger \psi_1 - \psi_2^\dagger \psi_2 \right) 
+ \frac{1}{2} \left( \partial_y A \right)^2 \Biggr]. \label{antipodal}
\eea
Here $g$ is the gauge coupling constant, and $A$ is the single component of the photon in $d=2$ which is transverse to ${\bf q}$.
This theory has been studied in great detail in recent work \cite{sslee,metnem,mross}, and it was found that the fermion
temporal derivative terms are irrelevant in the scaling limit. Here we will assume that this is the case, and show how this fixes the value of $z$. It is easy to see that the spatial gradient terms in $\mathcal{S}$ are invariant under the following scaling transformations:
\bea
&& x \rightarrow x/s, \quad y \rightarrow y/s^{1/2}, \quad \tau \rightarrow \tau/s^z, \nn
&& \quad  A \rightarrow  A \,  s^{(2z+1)/4}, \quad \psi \rightarrow  \psi \, s^{(2z+1)/4}. 
\eea
Then the gauge coupling constant in Eq.~(\ref{antipodal}) is found to transform as
\beq
g  \rightarrow  g \, s^{(3-2z)/4},
\eeq
and we see that a fixed point theory requires $z=3/2$ at tree level. The unusual feature of this computation is that we have used
the invariance of an interaction term to fix the value of $z$. Usually, $z$ is determined by demanding invariance of the temporal
derivative terms which are quadratic in the fields; however, such terms are strongly irrelevant here, and so can be set to zero at the outset.
Indeed the irrelevance of terms like $\psi^\dagger \partial_\tau \psi$ is an inevitable characteristic of a non-Fermi liquid, because then 
the dominant frequency dependence of the fermion Green's function arises from the self energy.
This opens the possibility of determining $z$ by fixing the strength of a boson-fermion interaction. 
In the present case, a lengthy computation\cite{metnem} with $\mathcal{S}$ 
shows that such a tree-level value of $z$ has no corrections to three loops.

\newpage
\section{Connections to string theory}
\label{sec:string}

Recent years have seen a significant effort to realize strongly-coupled conformal and compressible phases of matter
using the methods of gauge gravity duality.\cite{arcmp} Underlying this connection is the AdS/CFT correspondence which
provides a duality between CFTs in $d+1$ spacetime dimensions, and theories of gravity in $d+2$ dimensional anti-de Sitter 
space (AdS$_{d+2}$). 

\begin{figure}[htbp]
\begin{center}
 \includegraphics[width=4in]{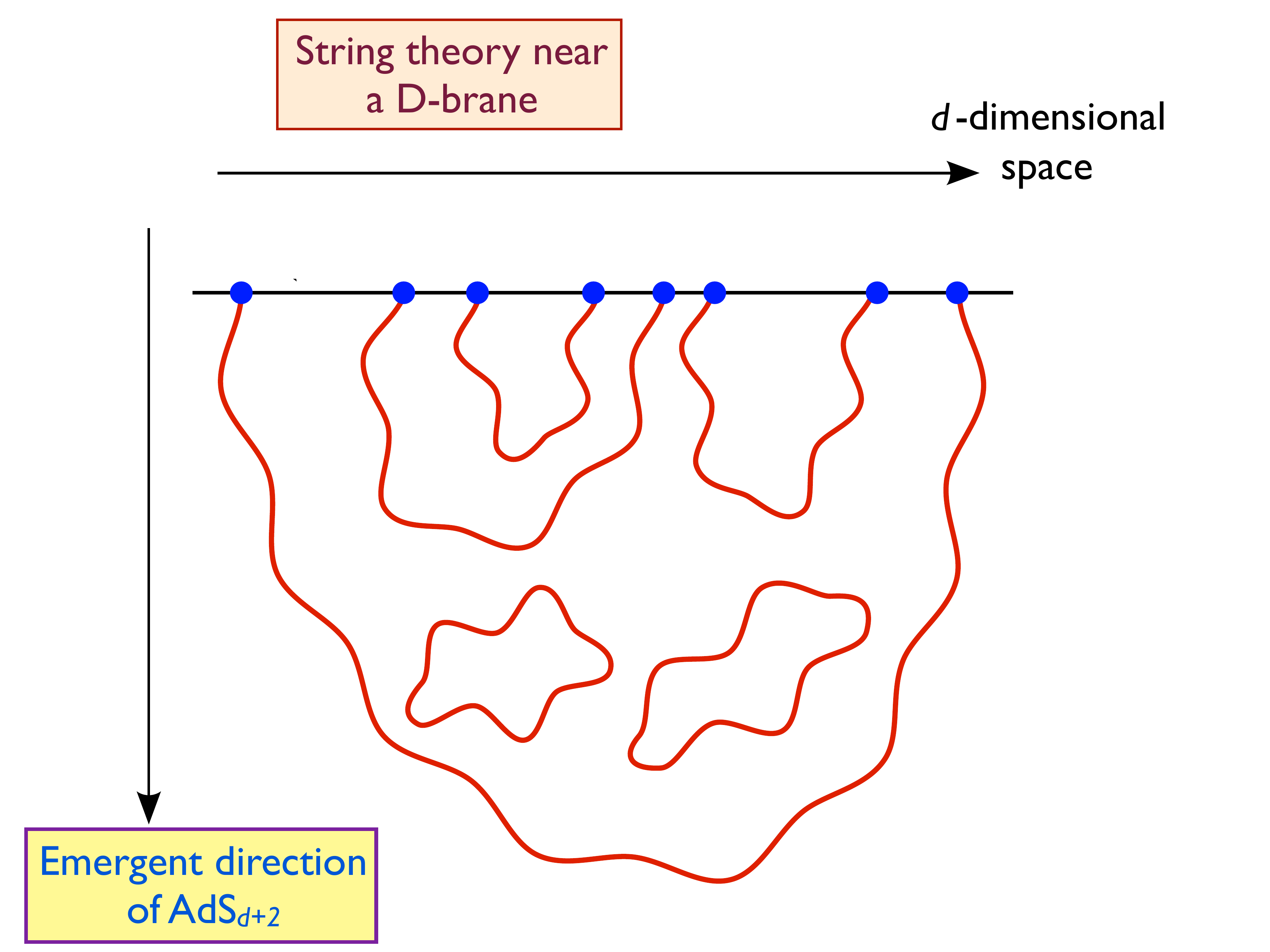}
 \caption{A $D$-brane in string theory. The strings end on a $d$-dimensional spatial surface. The blue circles represent the particles
 of quantum matter.}
\label{fig:strings}
\end{center}
\end{figure}

\begin{figure}[htbp]
\begin{center}
 \includegraphics[width=4in]{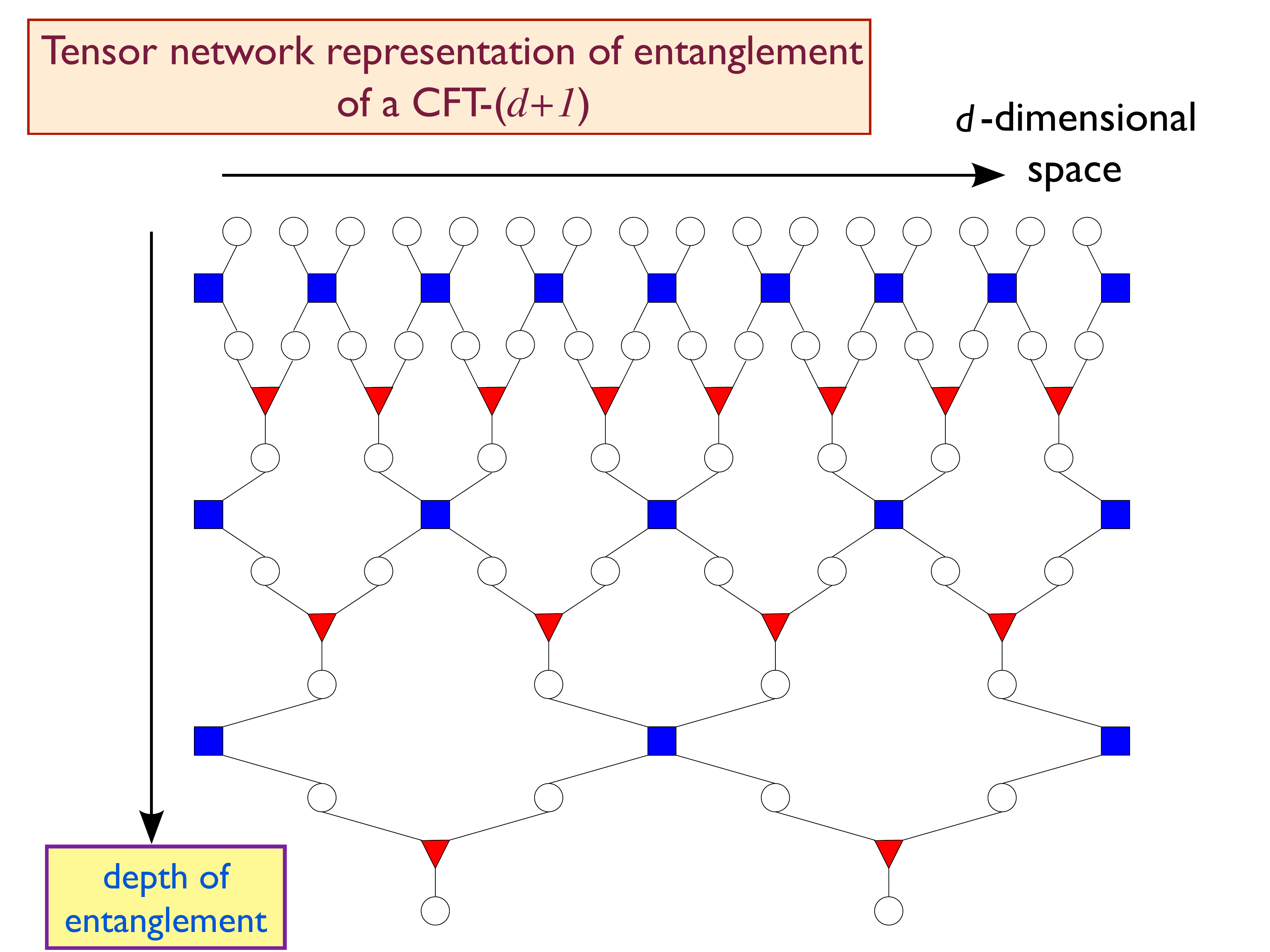}
 \caption{Pictorial tensor network representation (from Ref.~\refcite{swingle}) of entanglement on a lattice model of quantum degrees of freedom
represented by the open circles in the top row.}
\label{fig:tensor}
\end{center}
\end{figure}

\begin{figure}[htbp]
\begin{center}
 \includegraphics[width=4in]{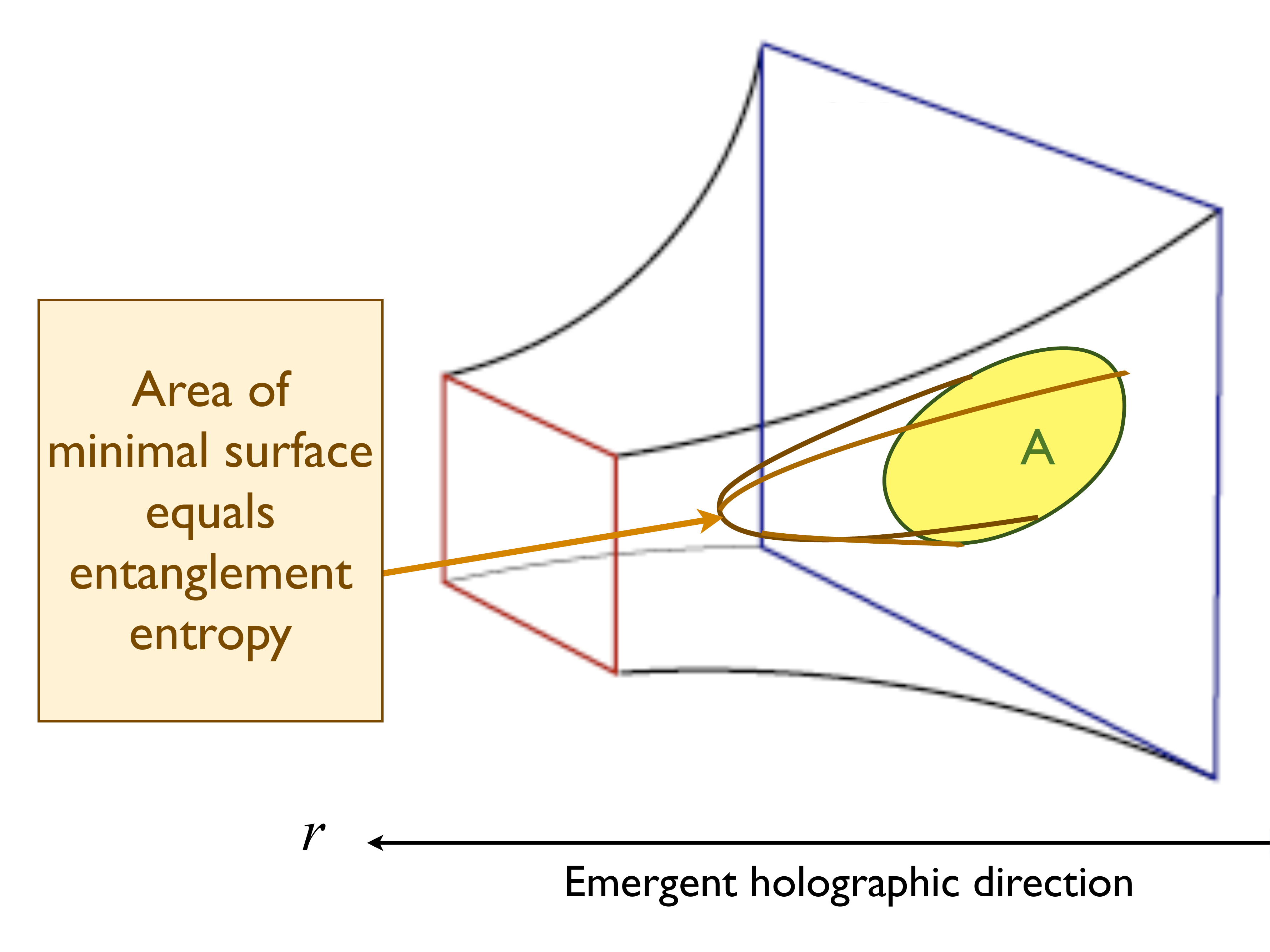}
 \caption{Computation of the entanglement entropy defined in Fig.~\ref{fig:ee} of region A. The Ryu-Takayanagi formula equates
 $S_E$ to the area of the minimal surface enclosing region A in the gravity theory.}
\label{fig:hee}
\end{center}
\end{figure}

An intuitive picture of the correspondence is provided by the picture of a $d$-dimensional $D$-brane of string theory 
shown in Fig.~\ref{fig:strings}. 
The low energy limit of the string theory is a CFT-$(d+1)$, representing the quantum matter we are interested in. 
The strings move in AdS$_{d+2}$, and can be seen as the source of long-range entanglement in the quantum matter.
This is highlighted by the similarity between Fig.~\ref{fig:strings} and the tensor network representation of 
entanglement\cite{levin,vidal} in Fig.~\ref{fig:tensor}.
In this connection, the emergent spatial direction of AdS$_{d+2}$ is seen to represent the depth of entanglement between the 
quantum matter degrees of freedom.\cite{swingle}
The Ryu-Takayanagi formula\cite{rt} for the entanglement entropy (Fig.~\ref{fig:hee}) also emerges from this connection via a computation
of the entanglement entropy from the tensor network.

The early applications of gauge-gravity duality to condensed matter physics\cite{m2} addressed issues related to 
the $T>0$ quantum-critical dynamics (Fig~\ref{fig:qc}) of conformal quantum matter; these have been reviewed recently elsewhere.\cite{arcmp} Here, I will briefly describe recent ideas on its application to compressible quantum matter.

Let us take as our objective the determination of the gravity dual of the theory in Eq.~(\ref{nfl}) describing 
non-zero density fermions coupled to a gauge field. As argued by 't Hooft,\cite{hooft} such duals are obtained in a suitable large $N$ limit.
In the condensed matter latter literature, the fermion $\psi$ is endowed with $N_f$ flavors and the large $N_f$ has been intensively examined.\cite{reizer,palee,monien,hlr,polchinski,nayak,bim,furusaki,metnem,sslee,mross,metzner,bartosch} 
At leading order in $1/N_f$ in $d=2$, computations from Eq.~(\ref{antipodal}) show 
that the fermion Green's function is modified from the Fermi liquid form
in Eq.~(\ref{fl}) by a singular correction which approaches the non-Fermi liquid scaling structure in Eq.~(\ref{scale}); schematically, this correction is of the form
\beq
G_\psi^{-1} \approx \omega - v_F q + i \frac{c}{N_f} \omega^{2/3}, \label{nf}
\eeq
which exhibits the $z=3/2$ scaling discussed below Eq.~(\ref{antipodal}).
Note that the term of Eq.~(\ref{nf}) which is 
most singular in the low energy limit has a prefactor of $1/N_f$. This is dangerous, and leads to a breakdown in the 
bare Feynman graph structure of the $1/N_f$ expansion\cite{sslee,metnem}; even at first order in $1/N_f$, it is necessary to at least sum
all planar graphs.

So let us consider an alternative case \cite{yamada,liza,hyper,tadashi2} 
where the gauge group is SU($N_c$), and we take the fermions $\psi$ to transform
under the {\em adjoint\/} representation of SU($N_c$). Then, an analysis of the Feynman graph expansion
shows that the low loop contributions to the $N_c \rightarrow \infty$ theory have the same low frequency structure as in Eq.~(\ref{nf})
in $d=2$, but {\em without\/} suppression of the singular terms by powers of $1/N_c$:
\beq
G_\psi^{-1} \approx \omega - v_F q + i \tilde{c} \, \omega^{2/3}. \label{nfmatrix}
\eeq
This indicates that the Feynman graph counting of powers of $1/N_c$ holds in the $N_c \rightarrow \infty$ limit,
and is identical to that in the classic paper by 't Hooft.\cite{hooft} Consequently, even at non-zero density and in the critical
low energy theory, the $1/N_c$ expansion is an expansion in powers 
of the genus of the surface defined by the double-line Feynman graphs. By the arguments of 't Hooft,\cite{hooft} we can reasonably hope
that the $N_c \rightarrow \infty$ theory is described by a dual gravity theory. Furthermore, given the issues with the 
$1/N_f$ expansion noted above, the $N_c \rightarrow \infty$ limit appears to be suited to capture the physics
of condensed matter systems.

Now, we will constrain the background metric of this hypothetical gravity theory by general scaling arguments.\cite{hyper}
We represent the $d$-dimensional spatial displacement by $\bm{dx}$, time displacement by $dt$, the emergent direction by $dr$,
and proper distance on the holographic space by $ds$. We are interested in states with a low energy scaling symmetry, and so
we demand that the low energy metric obey
\bea
\bm{x} &\rightarrow& \zeta \, \bm{x} \nn
t &\rightarrow& \zeta^z \, t \nn
ds &\rightarrow & \zeta^{\theta/d} \, ds . \label{scaling}
\eea
under rescaling by a factor $\zeta$. This defines $z$ as the dynamic critical exponent, and we now argue that $\theta$ is the violation
of hyperscaling exponent which was defined earlier by Eq.~(\ref{therm2}).
Using translation and rotational invariance in space, and translational invariance in time, we deduce the metric
\beq
ds^2 = \frac{1}{r^2} \left( - \frac{dt^2}{r^{2d (z-1)/(d-\theta)}} + r^{2\theta/(d-\theta)} dr^2 + \bm{d x}^2 \right), \label{zmetric}
\eeq
as the most general solution to (\ref{scaling}) modulo prefactors and reparametrization invariance in $r$. For our choice of the co-ordinates
in (\ref{zmetric}), $r$ transforms as
\beq
r \rightarrow \zeta^{(d-\theta)/d} r. \label{rscaling}
\eeq
Now let us take this gravity theory to a temperature $T>0$.
This thermal state requires a horizon, and let us assume the horizon appears at $r=r_H$. The entropy density
of this thermal state, $S$, will be proportional to the spatial area of the horizon, and so from Eq.~(\ref{zmetric})
we have $S \sim r_H^{-d}$. Now $T$ scales as $1/t$, and so from Eqs.~(\ref{scaling}) and (\ref{rscaling}) we deduce
$r_H^{-d} \sim T^{(d-\theta)/z}$ and so
$S \sim T^{(d-\theta)/z}$,
which matches the definition in Eq.~(\ref{therm2}). This justifies our identification of the $\theta$ appearing in Eq.~(\ref{zmetric})
as the violation
of hyperscaling exponent.\cite{hyper,tadashi1,hyper2,hyper3,hyper4,hyper5,hsingh,dey1,dey2}

With this gravitational definition of $z$ and $\theta$, we can now obtain additional properties
of these exponents which should also apply to the dual field theory. Remarkably, there is no known derivation of these properties
directly from the field theory.
We expect Eq.~(\ref{zmetric}) to be a solution of the analog of Einstein's equations in some gravitational theory; so it is reasonable
to impose the null energy condition\cite{tadashi1}, and this yields the important inequality\cite{hyper}
\beq
z \geq 1 + \frac{\theta}{d}. \label{null}
\eeq
As a final general property of the quantum matter which can be computed directly from the holographic metric in Eq.~(\ref{zmetric}), we turn to the entanglement entropy. This can be computed, via the Ryu-Takayanagi formula \cite{rt}, by computing the minimal surface area of Fig~\ref{fig:hee}, and we find \cite{tadashi1,hyper}
\beq
S_E \sim \left\{ \begin{array}{clc} 
P & , & \mbox{for $\theta < d-1$}\\
P \ln P & , & \mbox{for $\theta = d-1$}\\
P^{\theta/(d-1)} & , & \mbox{for $\theta > d-1$}
\end{array} \right. \label{Stheta}
\eeq
where $P$ is the surface area ({\em i.e.\/} the perimeter in $d=2$) of region A, as in Eq.~(\ref{see}).
Note that the `area law' of entanglement entropy is obeyed only for $\theta < d-1$.
The regime $\theta > d-1$ has strong violations of the area law, and so this is unlikely to represent a generic local quantum
field theory.

Note that here we defined $z$ and $\theta$ as exponents which appear in the metric of the gravitation theory in Eq.~(\ref{zmetric}).
However, as we have shown above, they also have independent definitions in terms of the boundary quantum theory via Eq.~(\ref{therm2}).
One of the important consequences of the gravitational definition is that we are now able to conclude that these exponents obey 
the inequality Eq.~(\ref{null}), and constrain the entanglement entropy as in Eq.~(\ref{Stheta}). No independent field-theoretic derivation
of these results is known. Indeed, Eqs.~(\ref{null}) and (\ref{Stheta}) may be taken as necessary conditions for 
the existence of a reasonable gravity dual of the field theory.

So far, our gravitational scaling analysis has been very general, and could apply to any dual critical theory. 
We now compare to the results of the field theoretic analysis discussed
in Section~\ref{sec:nfl} for the non-Fermi liquid state of fermions coupled to a gauge field. 
We use the temperature dependence of the thermal entropy to fix the value of $\theta=d-1$ found in Eq.~(\ref{theta}).
The combination of Eqs.~(\ref{zmetric}) and (\ref{theta}) is then the metric of the hypothetical gravitational dual \cite{tadashi1,hyper} 
description of strongly interacting compressible quantum matter, such as that realized by the 
$N_c \rightarrow \infty$ limit of the theory in Eq.~(\ref{nfl}) with an SU($N_c$) gauge field and fermions in the adjoint 
representation of SU($N_c$).

This proposal, and in particular the value of $\theta$ in Eq.~(\ref{theta}), can now be subjected to a number of tests:
\begin{itemize}
\item In $d=2$, we have $\theta = 1$ from Eq.~(\ref{theta}), and so from Eq.~(\ref{null}) the gravity dual theory requires $z \geq 3/2$. 
Remarkably, the lower bound, $z=3/2$, is the value obtained from the weak-coupling field theory analysis extended to 
three loops,\cite{metnem} as we discussed below Eq.~(\ref{antipodal}).
\item We see from Eq.~(\ref{Stheta}) that
for the value of $\theta$ in Eq.~(\ref{theta}), there is logarithmic violation of the area law,\cite{tadashi1} as expected
for a system with a Fermi surface.\cite{tarun1,swingleprl}
\item The metric in Eq.~(\ref{zmetric}) appears as a solution\cite{kiritsis,trivedi,kiritsis2} 
of a class of Einstein-Maxwell-dilaton theories. In this realization, 
there is a non-zero charge density $\langle \mathcal{Q}\rangle$ on the $d$-dimensional boundary, and the 
compressibility $d\langle \mathcal{Q}\rangle/d\mu$ is non-zero.
\item
A complete computation of the entanglement entropy in the Einstein-Maxwell-dilaton theory yields the following
expression for the entanglement entropy\cite{hyper}
\beq
S_E = \lambda \langle \mathcal{Q}\rangle^{(d-1)/d} P \ln P, \label{seef}
\eeq
A key feature\cite{hyper} is that the dependence upon the shape of region
A is only through the value of $P$, and the prefactor $\lambda$ is {\em independent\/} of the shape or any other geometric of 
property of region A: this matches the characteristics of 
the entanglement entropy of a spherical Fermi surface.\cite{swingleprl,klich} 
\item
The value of $\lambda$ in Eq.~(\ref{seef}), by a variant of the attractor mechanism \cite{andy,sen,prakash}, 
is independent of all ultraviolet details
of the gravity theory \cite{hyper}, and $S_E$ depends only upon the value of $\langle\mathcal{Q}\rangle$ as shown.
This supports the conclusion that the prefactor of the entanglement entropy of a non-Fermi liquid is universal,
as it is believed to be for an interacting Fermi liquid.\cite{swingleprl}
\item We expect a Luttinger relation for the volume of the `hidden' Fermi surface, with $\langle \mathcal{Q} \rangle \sim k_F^d$.
Then the $k_F$ dependence of Eq.~(\ref{seef}) is that expected for a Fermi surface: this can be viewed as indirect evident
for the Luttinger relation.\cite{hyper} In this manner, the Luttinger relation, which is one of the deepest results of 
condensed matter physics, is surprisingly connected 
to two fundamental features of the holographic theory: Gauss' Law and the attractor mechanism \cite{hyper,ssfl}.
\item Refs.~\refcite{hyper,lizasean} also studied the transition of the 
Einstein-Maxwell-dilaton theory to a state with {\em partial\/} confinement,
in which there were additional Fermi surfaces of gauge-neutral particles. The resulting state is analogous to the `fractionalized Fermi liquid'
of Kondo and Hubbard models.\cite{ffl1,sces} It was found\cite{hyper} that the holographic entanglement entropy of this partially 
confined state was given by Eq.~(\ref{seef}) but with $\langle\mathcal{Q}\rangle \rightarrow \langle\mathcal{Q} - 
\mathcal{Q}_{\rm conf}\rangle$, where $\langle \mathcal{Q}_{\rm conf} \rangle$ is the density associated with the Fermi surfaces
of gauge-neutral particles. If we now use Eq.~(\ref{seef}) to fix the $k_F$ of the hidden Fermi surfaces of gauge-dependent 
particles just as above, then we see that the Luttinger relation for $\langle \mathcal{Q} \rangle$ equates it to the sum
of the gauge-neutral and gauge-charged Fermi surfaces, again as expected from the gauge theory analysis.\cite{ffl1}
\end{itemize}
Clearly, it would be useful to ultimately obtain evidence for the wavevector $k_F$ of the hidden Fermi surfaces
by the spatial modulation of some response function. Short of such confirmation, the above tests do provide strong evidence 
for the presence of a hidden Fermi surface in such gravity theories of compressible quantum matter. These gravity theories appear as 
solutions of the Einstein-Maxwell-dilaton theories\cite{kiritsis,trivedi,kiritsis2} which contain only bosonic degrees of freedom; so they may be 
viewed as analogs of the `bosonization' of the Fermi surface.\cite{neto1,neto2,haldane,marston,lawler1,lawler2,senthilbose}
\newpage
\section*{Acknowledgments} 

I am very grateful to D.~Chowdhury, P.~Fendley, E.~Fradkin, S.~Kachru, L.~Huijse, M.~Metlitski, R.~Nandkishore, and D.~Vegh for valuable comments. I would like to especially thank S.~Kivelson for detailed comments on all aspects of the manuscript.
This research was supported by the National Science Foundation under grant DMR-1103860, and by a MURI grant from AFOSR.


\begin{thebibliography}{99}

\bibitem{epr} A. Einstein, B. Podolsky, and N. Rosen, ``Can Quantum-Mechanical Description of Physical Reality Be Considered 
Complete?'',
Phys. Rev. {\bf 47}, 777 (1935).

\bibitem{bell} J.~S.~Bell, ``On the Einstein Podolsky Rosen paradox,'' Physics {\bf 1}, 195 (1964).

\bibitem{cherny} S.~R.~White and A.~L.~Chernyshev, ``Neel order in square and triangular lattice Heisenberg models,''
Phys. Rev. Lett. {\bf 99}, 127004 (2007). [arXiv:0705.2746]

\bibitem{fazekas} P. Fazekas and P. W. Anderson, ``On the ground state properties of the anisotropic triangular antiferromagnet,''
Philos. Mag. {\bf 30}, 423 (1974).

\bibitem{pauling} L.~Pauling, ``A Resonating-Valence-Bond Theory of Metals and Intermetallic Compounds,''
Proc. Roy. Soc. London A {\bf 196}, 343 (1949).

\bibitem{pwa} P. W. Anderson, ``The resonating valence bond state in La$_2$CuO$_4$ and superconductivity,'' Science {\bf 235}, 1196 (1987).

\bibitem{KRS} S.~A.~Kivelson, D.~S.~Rokhsar, and J.~P.~Sethna,
``Topology of the resonating valence-bond state: Solitons and high-$T_c$ superconductivity,''
Phys. Rev. B {\bf 35}, 8865 (1987).

\bibitem{bza} G. Baskaran and P. W. Anderson, ``Gauge theory of high-temperature superconductors and 
strongly correlated Fermi systems,''
Phys. Rev. B {\bf 37}, 580 (1988).

\bibitem{rokhsar} D.~Rokhsar and S.~A.~Kivelson, ``Superconductivity and the Quantum Hard-Core Dimer Gas,''
Phys. Rev. Lett. {\bf 61}, 2376 (1988).

\bibitem{fradkiv} E.~Fradkin, ``The Spectrum of Short-Range Resonating Valence Bond Theories,'' 
in {\em Field Theories in Condensed Matter Physics: A Workshop}, edited by Z. Tesanovic, Addison-Wesley (1990);
E.~Fradkin and S.~A.~Kivelson, ``Short range resonating valence bond theories and superconductivity,''
Mod. Phys. Lett. B {\bf 4}, 225 (1990).

\bibitem{rsb} N. Read and S. Sachdev, ``Spin-Peierls, valence bond solid, and Neel ground states of low dimensional quantum antiferromagnets,''  Phys. Rev. B {\bf 42}, 4568 (1990). 

\bibitem{laughlin} V.~Kalmeyer and R.~B.~Laughlin, ``Equivalence of the resonating-valence-bond and fractional quantum Hall states,''
Phys. Rev. Lett. {\bf 59},
2095 (1987).

\bibitem{wwz} X.-G. Wen, F. Wilczek, and A. Zee, ``Chiral Spin States and Superconductivity,'' Phys. Rev. B {\bf 39},
11413 (1989).

\bibitem{wen1} X.-G.~Wen, ``Mean-field theory of spin-liquid states with finite energy gap and topological orders,''
Phys. Rev. B {\bf 44}, 2664 (1991).

\bibitem{rstl} N. Read and S. Sachdev, ``Large-$N$ expansion for frustrated quantum antiferromagnets,'' Phys. Rev. Lett. {\bf 66}, 1773 (1991).

\bibitem{sr} S. Sachdev and N. Read, ``Large $N$ 
expansion for frustrated and doped quantum antiferromagnets,'' Int. J. Mod. Phys. B {\bf 5}, 219 (1991) [arXiv:cond-mat/0402109].

\bibitem{jalabert} R. Jalabert and S. Sachdev, ``Spontaneous alignment of frustrated bonds in an anisotropic, three dimensional Ising model,''
Phys. Rev. B {\bf 44}, 686 (1991). This paper describes the dual form of the gauge theory of the $\mathbb{Z}_2$ RVB phase; details of the derivation 
were published in S.~Sachdev
and M.~Vojta, ``Translational symmetry breaking in two-dimensional antiferromagnets and superconductors,''
J. Phys. Soc. Japan {\bf 69}, Suppl. B, 1 (2000) [arXiv:cond-mat/9910231].

\bibitem{sf} T.~Senthil and M.~P.~A.~Fisher, ``$\mathbb{Z}_2$ Gauge Theory of Electron 
Fractionalization in Strongly Correlated Systems,'' 
Phys. Rev. B {\bf 62}, 7850 (2000). [arXiv:cond-mat/9910224]

\bibitem{sondhi} R.~Moessner, S.~L.~Sondhi, and E.~Fradkin, ``Short-ranged RVB physics, quantum dimer models and 
Ising gauge theories,''
Phys. Rev. B {\bf 65}, 024504 (2001). [arXiv:cond-mat/0103396]

\bibitem{fendley} E.~Ardonne, P.~Fendley, and E.~Fradkin,
``Topological Order and Conformal Quantum Critical Points,''
Annals of Physics {\bf 310}, 493 (2004).
[arXiv:cond-mat/0311466]

\bibitem{sstri} S.~Sachdev,
``Kagome and triangular lattice Heisenberg antiferromagnets: ordering from quantum fluctuations and quantum-disordered
ground states with deconfined bosonic spinons'',
Phys. Rev. B {\bf 45}, 12377 (1992). This paper discusses $\mathbb{Z}_2$ RVB phases on the kagome and triangular lattices
in general, and notes that these should appear in the quantum dimer model on these lattices (page 12393).

\bibitem{moessner} R.~Moessner and S.~L.~Sondhi, ``Resonating Valence Bond Phase in the Triangular Lattice Quantum Dimer Model,''
Phys. Rev. Lett. {\bf 86}, 1881 (2001). [arXiv:cond-mat/0007378]

\bibitem{pasquier}  G.~Misguich, D.~Serban, and V.~Pasquier, ``Quantum dimer model on the kagome lattice: solvable dimer liquid and Ising gauge theory,'' Phys. Rev. Lett. {\bf 89}, 137202 (2002). [arXiv:cond-mat/0204428]

\bibitem{yao} Hong Yao and S.~A.~Kivelson,
``Exact Spin Liquid Ground States of the Quantum Dimer Model on the Square and Honeycomb Lattices,''
	arXiv:1112.1702.

\bibitem{thouless} D.~J.~Thouless, ``Fluxoid quantization in the resonating-valence-bond model,''
Phys. Rev. B {\bf 36}, 7187 (1987).

\bibitem{rc} N.~Read and B.~Chakraborty, ``Statistics of the excitations of the resonating-valence-bond state,''
Phys. Rev. B {\bf 40}, 7133 (1989).

\bibitem{kitaev1} A.~Y.~Kitaev,
``Fault-tolerant quantum computation by anyons,''
Annals of Physics  {\bf 303}, 2 (2003). [arXiv:quant-ph/9707021]

\bibitem{kitaev2} A.~Y.~Kitaev,
``Anyons in an exactly solved model and beyond,''
Annals of Physics  {\bf 321}, 2 (2006). [arXiv:cond-mat/0506438]

\bibitem{tarun2} T.~Grover, A.~M.~Turner, and A.~Vishwanath, 
``Entanglement Entropy of Gapped Phases and Topological Order in Three dimensions,''
Phys. Rev. B {\bf 84}, 195120 (2011). [arXiv:1108.4038]

\bibitem{preskill} A. Kitaev and J. Preskill, ``Topological entanglement entropy,''
Phys. Rev. Lett. {\bf 96}, 110404 (2006). [arXiv:hep-th/0510092]

\bibitem{levinwen} M. Levin and X.-G. Wen, ``Detecting topological order in a ground state wave function,''
Phys. Rev. Lett. {\bf 96}, 110405 (2006). [arXiv:cond-mat/0510613]

\bibitem{tarun1} Y.~Zhang, T. Grover, and A. Vishwanath, ``Topological Entanglement Entropy of $\mathbb{Z}_2$ 
Spin liquids and Lattice Laughlin states,''
Phys. Rev. B  {\bf 84}, 075128 (2011). [arXiv:1106.0015]

\bibitem{bais1} F.~A.~Bais, ``Flux metamorphosis,'' 
Nucl. Phys. B {\bf 170}, 3243 (1980). 

\bibitem{bais2} F.~A.~Bais, P.~van Driel, and M.~de Wild Propitius, ``Quantum symmetries in discrete gauge
theories,'' Phys. Lett. B {\bf 280}, 63 (1992).

\bibitem{seiberg}  J.~M.~Maldacena, G.~W.~Moore and N.~Seiberg,
  ``D-brane charges in five-brane backgrounds,''
  JHEP {\bf 0110}, 005 (2001). [arXiv:hep-th/0108152]

\bibitem{freedman} M. Freedman, C. Nayak, K. Shtengel, K. Walker, and 
Z. Wang, ``A Class of $P,T$-Invariant Topological Phases of Interacting Electrons,''
Annals of Physics {\bf 310}, 428 (2004). [arXiv:cond-mat/0307511]

\bibitem{cenke} C.~Xu and S.~Sachdev, ``Global phase diagrams of frustrated quantum antiferromagnets in two dimensions: 
doubled Chern-Simons theory,''
Phys. Rev. B {\bf 79}, 064405 (2009). [arXiv:0811.1220]

\bibitem{flammia} S. T. Flammia, A. Hamma, T. L. Hughes, and X.-G. Wen, ``Topological Entanglement Renyi Entropy and Reduced Density Matrix Structure,''
Phys. Rev. Lett. {\bf 103}, 261601 (2009). [arXiv:0909.3305]

\bibitem{igor1}   I.~R.~Klebanov, S.~S.~Pufu, S.~Sachdev and B.~R.~Safdi,
  ``Renyi Entropies for Free Field Theories,''
  arXiv:1111.6290.
  
\bibitem{sskagome} S.~Sachdev,
``Kagome and triangular lattice Heisenberg antiferromagnets: ordering from quantum fluctuations and quantum-disordered
 ground states with deconfined bosonic spinons'',
Phys. Rev. B {\bf 45}, 12377 (1992).

\bibitem{whitekagome} Simeng Yan, D.~A.~Huse, and S.~R.~White, 
``Spin Liquid Ground State of the $S=1/2$ Kagome Heisenberg Model,'' Science {\bf 332}, 1173 (2011). [arXiv:1011.6114]

\bibitem{punk} Y.~Huh, M.~Punk and S.~Sachdev, ``Vison states and confinement transitions of $\mathbb{Z}_2$ 
spin liquids on the kagome lattice,''
Phys. Rev. B {\bf 84}, 094419 (2011). [arXiv:1106.3330]

\bibitem{messio}  L.~Messio, B.~Bernu, and C.~Lhuillier, ``The kagome antiferromagnet: a chiral topological spin liquid ?''
arXiv:1110.5440.

\bibitem{log2} Hong-Chen Jiang, Zhenghan Wang, and L.~Balents, ``Identifying Topological Order by Entanglement Entropy,''
 arXiv:1205.4289.

\bibitem{younglee} Young Lee, ``Experimental signatures of spin liquid physics on the $S=1/2$ kagome lattice,'' 
Bull. Am. Phys. Soc. {\bf 57}, 1, H8.00005 (2012).

\bibitem{bfg} L.~Balents, M.~P.~A.~Fisher, and S.~M.~Girvin, ``Fractionalization in an Easy-axis Kagome Antiferromagnet,''
Phys. Rev. B {\bf 65}, 224412 (2002). [arXiv:cond-mat/0110005]

\bibitem{isakov1} S.~V.~Isakov, Y.-B.~Kim, and A.~Paramekanti, 
``Spin liquid phase in a $S=1/2$ quantum magnet on the kagome lattice,'' Phys. Rev. Lett. {\bf 97}, 207204 (2006). [arXiv:cond-mat/0607778]

\bibitem{isakov2} S.~V.~Isakov, M.~B.~Hastings, and R.~G.~Melko,
``Topological Entanglement Entropy of a Bose-Hubbard Spin Liquid,''
Nature Physics {\bf 7}, 772 (2011). [arXiv:1102.1721]

\bibitem{j1j20} F. Figueirido,
A. Karlhede, 
S. Kivelson, S. Sondhi, 
M. Rocek, and 
D. S. Rokhsar, ``Exact diagonalization of finite frustrated spin-1/2 Heisenberg models,'' Phys. Rev. B {\bf 41}, 4619 (1990).

\bibitem{j1j21} Hong-Chen Jiang, Hong Yao, and L.~Balents, ``Spin Liquid Ground State of the 
Spin-1/2 Square $J_1$-$J_2$ Heisenberg Model,''  arXiv:1112.2241.

\bibitem{j1j22} Ling Wang, Zheng-Cheng Gu, Xiao-Gang Wen, and F.~Verstraete,
``Possible spin liquid state in the spin 1/2 $J_1$-$J_2$ antiferromagnetic Heisenberg model on square lattice: A tensor product state approach,''
arXiv:1112.3331.

\bibitem{sorella}  Tao Li, F.~Becca, Wenjun Hu, and S,~Sorella, 
``Gapped spin liquid phase in the $J_1$-$J_2$ Heisenberg model by a Bosonic resonating valence-bond ansatz,''
arXiv:1205.3838.

\bibitem{janke}  S.~Wenzel and W.~Janke, ``Comprehensive quantum Monte Carlo study of the quantum critical points in 
planar dimerized/quadrumerized Heisenberg models,''
Phys. Rev. B {\bf 79}, 014410 (2009). [arXiv:0808.1418]

\bibitem{mfs} M.~A.~Metlitski, C.~A.~Fuertes, and S.~Sachdev, 
``Entanglement Entropy in the O($N$) model,'' 
Phys.\ Rev.\ B {\bf 80}, 115122 (2009). 
[arXiv:0904.4477]. 

\bibitem{brown} D.~S.~Chow, P.~Wzietek, D.~Fogliatti, B.~Alavi, D.~J.~Tantillo, C.~A.~Merlic, and S.~E.~Brown, 
``Singular Behavior in the Pressure-Tuned Competition between Spin-Peierls and Antiferromagnetic Ground States of (TMTTF)$_2$PF$_6$,''
Phys. Rev. Lett. {\bf 81}, 3984 (1998).

\bibitem{ruegg} Ch.~Ruegg,  B. Normand, M. Matsumoto, A. Furrer, 
D. F. McMorrow, K. W. Kr\"amer, H. -U. G\"udel, S. N. Gvasaliya, H. Mutka, and M. Boehm,  
``Quantum Magnets under Pressure: Controlling Elementary Excitations in TlCuCl$_3$,'' Phys. Rev. Lett. {\bf 100}, 205701 (2008).
[arXiv:0803.3720]

\bibitem{normand} B.~Normand and T.~M.~Rice, ``Dynamical Properties of an Antiferromagnet near the Quantum Critical Point: Application 
to LaCuO$_{2.5}$,''
Phys. Rev. B {\bf 56}, 8760 (1997).
[arXiv:cond-mat/9701202]

\bibitem{crossover}S. Sachdev, `` Theory of finite temperature crossovers near quantum critical points close to, or above, their 
upper-critical dimension,'' Phys. Rev. B {\bf 55}, 142 (1997). [arXiv:cond-mat/9606083]

\bibitem{solvay08} S.~Sachdev, `` Exotic phases and quantum phase transitions: model systems and experiments,''
Rapporteur presentation at the 24th Solvay Conference on 
Physics, ``Quantum Theory of Condensed Matter'', Brussels, Oct 11-13, 2008, arXiv:0901.4103.

\bibitem{rsl} N.~Read and S.~Sachdev, ``Valence bond and spin-Peierls ground states of low dimensional quantum antiferromagnets,''
Phys. Rev. Lett. {\bf 62}, 1694 (1989).

\bibitem{senthil1} T.~Senthil, A.~Vishwanath, L.~Balents, S.~Sachdev, and
M.~P.~A.~Fisher, ``Deconfined quantum critical points,'' Science {\bf 303}, 1490 (2004). [cond-mat/0311326]

\bibitem{senthil2} T.~Senthil,
L.~Balents, S.~Sachdev, A.~Vishwanath, and M.~P.~A.~Fisher, `` Quantum criticality beyond 
the Landau-Ginzburg-Wilson paradigm,''
Phys.
Rev. B  {\bf 70}, 144407 (2004). [cond-mat/0312617]

\bibitem{sandvik} A.W. Sandvik, ``Evidence for deconfined quantum criticality in a two-dimensional Heisenberg model with four-spin interactions,''
Phys. Rev. Lett. {\bf 98}, 227202 (2007). [arXiv:cond-mat/0611343]

\bibitem{ribhu} R.~K.~Kaul and A.~W.~Sandvik,  ``A lattice model for the SU($N$) Neel-VBS quantum phase transition at large $N$,''
arXiv:1110.4130

\bibitem{sandvik2} A. W. Sandvik, ``Continuous quantum phase transition between an antiferromagnet and a valence-bond-solid 
in two dimensions; evidence for logarithmic corrections to scaling,''
Phys. Rev. Lett. {\bf 104}, 177201 (2010). [arXiv:1001.4296]

\bibitem{damle2} A. Banerjee, K. Damle, and F. Alet, ``Impurity spin texture at a deconfined quantum critical point,''
Phys. Rev. B {\bf 82},
155139 (2010). [arXiv:1002.1375]

\bibitem{bloch} M. Greiner, O. Mandel, T. Esslinger, T. W. H\"ansch, and I. Bloch, 
``Quantum phase transition from a superfluid to a Mott insulator in a gas of ultracold atoms,'' 
Nature {\bf 415}, 39 (2002).

\bibitem{fwgf} M.~P.~A.~Fisher, P.~B.~Weichman, G.~Grinstein, and D.~S.~Fisher,
``Boson localization and the superfluid-insulator transition,''
Phys.\ Rev.\ B {\bf 40}, 546 (1989).

\bibitem{chin} Xibo Zhang, Chen-Lung Hung, Shih-Kuang Tung, and Cheng Chin,
``Quantum critical behavior of ultracold atoms in two-dimensional optical lattices,''
Science {\bf 335}, 1070 (2012). [arXiv:1109.0344]

\bibitem{simon} J.~Simon, W.~S.~Bakr, Ruichao Ma, M.~E.~Tai, P.~M.~Preiss, and M.~Greiner,
``Quantum Simulation of Antiferromagnetic Spin Chains in an Optical Lattice,''
Nature {\bf 472}, 307 (2011). [arXiv:1103.1372]

\bibitem{kuhr} C.~Weitenberg, M.~Endres, J.~F.~Sherson, M.~Cheneau, P.~Schau{\ss}, T.~Fukuhara, 
I.~Bloch, and S.~Kuhr, ``Single-Spin Addressing in an Atomic Mott Insulator,''
Nature {\bf 471}, 319 (2011). [arXiv:1101.2076]

\bibitem{chn} S. Chakravarty, B.I. Halperin, and D.R. Nelson, 
``Low-temperature behavior of two-dimensional quantum antiferromagnets,''
Phys. Rev. Lett. {\bf 60}, 1057 (1988).

\bibitem{jinwu1} S. Sachdev and 
J. Ye, ``Universal quantum critical dynamics of two-dimensional antiferromagnets,''
Phys. Rev. Lett. {\bf 69}, 2411 (1992). [cond-mat/9204001]

\bibitem{jinwu2} A.~V.~Chubukov, S.~Sachdev, and J.~Ye, ``Theory of Two-Dimensional Quantum 
Heisenberg Antiferromagnets with a Nearly Critical Ground State,''
Phys. Rev. B {\bf 49}, 11919
(1994). [cond-mat/9304046]

\bibitem{arcmp} S.~Sachdev,  ``What can gauge-gravity duality teach us about condensed matter physics?''
Annual Review of Condensed Matter Physics
{\bf 3}, 9 (2012). [arXiv:1108.1197]

\bibitem{liza} L.~Huijse and S.~Sachdev, 
``Fermi surfaces and gauge-gravity duality,''
Phys. Rev. D {\bf 84}, 026001 (2011). [arXiv:1104.5022]  

\bibitem{yrz}  T. M. Rice, Kai-Yu Yang, and F.~C.~Zhang, 
``A Phenomenological Theory of the Anomalous Pseudogap Phase in Underdoped Cuprates,''
 Rep. Prog. Phys. {\bf 75}, 016502 (2012). [arXiv:1109.0632]

\bibitem{sces} S. Sachdev, M. A. Metlitski, and M. Punk, ``Antiferromagnetism in metals:
 from the cuprate superconductors to the heavy fermion materials,''
Proceedings of SCES 2011; arXiv:1202.4760. 

\bibitem{etsb} Minoru Yamashita, Norihito Nakata, Yoshinori Senshu, Masaki Nagata, Hiroshi M. Yamamoto,
Reizo Kato, Takasada Shibauchi, and Yuji Matsuda, 
``Highly Mobile Gapless Excitations in a Two-Dimensional Candidate Quantum Spin Liquid,''
Science {\bf 328}, 1246 (2010).

\bibitem{mot} O. I. Motrunich, 
``Variational study of triangular lattice spin-1/2 model with ring exchanges and spin liquid state 
in $\kappa$-(ET)$_2$Cu$_2$(CN)$_3$,''
Phys. Rev. B {\bf 72}, 045105 (2005).

\bibitem{mot2} Sung-Sik Lee and Patrick A. Lee, ``U(1) Gauge Theory of the Hubbard Model : Spin Liquid States and Possible Application 
to $\kappa$-(BEDT-TTF)$_2$Cu$_2$(CN)$_3$,''
Phys. Rev. Lett. {\bf 95}, 036403 (2005). [arXiv:cond-mat/0502139]

\bibitem{mot3} T.~Grover, N. Trivedi, T. Senthil, and Patrick A. Lee, ``Weak Mott insulators on the triangular lattice: possibility of a gapless nematic quantum spin liquid,''
Phys. Rev. B {\bf 81}, 245121 (2010). [arXiv:0907.1710]

\bibitem{biswas} R. R. Biswas, Liang Fu, C. Laumann, and S. Sachdev, ``SU(2)-invariant spin liquids on the triangular lattice 
with spinful Majorana excitations,''
Phys. Rev. B {\bf 83}, 245131 (2011). [arXiv:1102.3690]

\bibitem{lai} Hsin-Hua Lai and O.~I.~Motrunich,
``SU(2)-invariant Majorana spin liquid with stable parton Fermi surfaces in an exactly solvable model,''
Phys. Rev. B 84, 085141 (2011). [arXiv:1106.0028]

\bibitem{hermelemajorana} G.~Chen, A.~Essin, and M.~Hermele,
``Majorana spin liquids and projective realization of SU(2) spin symmetry,''
Phys. Rev. B {\bf 85}, 094418 (2012). [arXiv:1112.0586]

\bibitem{reizer} M.~Yu~Reizer,  ``Effective electron-electron interaction in metals and superconductors,''
Phys. Rev. B {\bf 39}, 1602 (1989).

\bibitem{palee} P.~A.~Lee, ``Gauge Field, Aharonov-Bohm Flux, and High-$T_c$ Superconductivity,'' Phys. Rev. Lett.
{\bf 63}, 680 (1989).

\bibitem{monien} B.~Blok and H.~Monien, 
``Gauge theories of high-$T_c$ superconductors,''
Phys. Rev. B {\bf 47}, 3454 (1993).

\bibitem{hlr} B. I. Halperin, P. A. Lee and N. Read, ``Theory of the half-filled Landau level,''
Phys. Rev. B {\bf 47}, 7312 (1993).

\bibitem{polchinski} J. Polchinski, ``Low Energy Dynamics of the Spinon-Gauge System,''
Nucl. Phys. B {\bf 422}, 617 (1994). [arXiv:cond-mat/9303037]

\bibitem{nayak} C. Nayak and F. Wilczek, ``Non-Fermi Liquid Fixed Point in 2+1 Dimension,'' 
Nucl. Phys. B 417, 359 (1994). [arXiv:cond-mat/9312086]

\bibitem{bim} B. L. Altshuler, L. B. Ioffe and A. J. Millis, ``On the low energy properies of fermions with singular interactions,''
Phys. Rev. B 50, 14048 (1994). [arXiv:cond-mat/9406024]
  
\bibitem{furusaki} Y.-B. Kim, A.~Furusaki, X.-G. Wen and P.~A.~Lee, 
``Gauge-invariant response functions of fermions coupled to a gauge field,''
Phys. Rev. B {\bf 50}, 17917 (1994). [arXiv:cond-mat//9405083]

\bibitem{sslee} Sung-Sik Lee, 
``Low energy effective theory of Fermi surface coupled with U(1) gauge field in 2+1 dimensions,'' 
Phys. Rev. B {\bf 80}, 165102 (2009). [arXiv:0905.4532]

\bibitem{metnem}  M.~A.~Metlitski, and S.~Sachdev,
``Quantum phase transitions of metals in two spatial dimensions: I. Ising-nematic order,''
  Phys.\ Rev.\  B {\bf 82}, 075127 (2010).
  [arXiv:1001.1153]

\bibitem{mross} D.~F.~Mross, J.~McGreevy, H.~Liu, and T.~Senthil, 
 ``A controlled expansion for certain non-Fermi liquid metals,''
  Phys. Rev. B {\bf 82}, 045121 (2010).
  [arXiv:1003.0894].

\bibitem{metzner} S.~C.~Thier and W.~Metzner, 
``Singular order parameter interaction at the nematic quantum critical point in two-dimensional electron systems,''
Phys. Rev. B 84, 155133 (2011). [arXiv:1108.1929]

\bibitem{bartosch} C.~Drukier, L.~Bartosch, A.~Isidori, and P.~Kopietz, 
``Functional renormalization group approach to the Ising-nematic quantum critical point of two-dimensional metals,''
arXiv:1203.2645.
  
\bibitem{levin} M. Levin and C. P. Nave, ``Tensor renormalization group approach to 2D classical lattice models,''
Phys. Rev. Lett. {\bf 99}, 120601 (2007). [arXiv:cond-mat/0611687]

\bibitem{vidal} G. Vidal, ``Entanglement renormalization,''
Phys. Rev. Lett. {\bf 99}, 220405 (2007).  [arXiv:cond-mat/0512165]

\bibitem{swingle} B.~Swingle, ``Entanglement Renormalization and Holography,''
arXiv:0905.1317.

\bibitem{rt} S.~Ryu and T.~Takayanagi,
 ``Holographic derivation of entanglement entropy from AdS/CFT,''
  Phys.\ Rev.\ Lett.\  {\bf 96}, 181602 (2006).
  [arXiv:hep-th/0603001]

\bibitem{m2} C. P. Herzog, P. Kovtun, S. Sachdev, and D. T. Son, 
``Quantum critical transport, duality, and M-theory,'' Phys. Rev. D {\bf 75}, 085020 (2007)
[arXiv:hep-th/0701036]

\bibitem{hooft} G.~'t Hooft, ``A Planar Diagram Theory for Strong Interactions,''
Nucl. Phys. B {\bf 72}, 461 (1974).


\bibitem{yamada}   D.~Yamada and L.~G.~Yaffe,
  ``Phase diagram of $\mathcal{N}$=4 super-Yang-Mills theory with R-symmetry chemical potentials,''
  JHEP {\bf 0609}, 027 (2006).
  [hep-th/0602074]

\bibitem{hyper}  L. Huijse, S. Sachdev, and B. Swingle, 
``Hidden Fermi surfaces in compressible states of gauge-gravity duality,''
Phys. Rev. B {\bf 85}, 035121 (2012). [arXiv:1112.0573]

\bibitem{tadashi2}   J.~Bhattacharya, N.~Ogawa, T.~Takayanagi and T.~Ugajin,
  ``Soliton Stars as Holographic Confined Fermi Liquids,''
  JHEP {\bf 1202}, 137 (2012).
  [arXiv:1201.0764]

\bibitem{tadashi1}   N.~Ogawa, T.~Takayanagi and T.~Ugajin,
  ``Holographic Fermi Surfaces and Entanglement Entropy,''
  JHEP {\bf 1201}, 125 (2012).
  [arXiv:1111.1023]
  
\bibitem{hyper2}   E.~Shaghoulian,
  ``Holographic Entanglement Entropy and Fermi Surfaces,''
  arXiv:1112.2702.

\bibitem{hyper3}   X.~Dong, S.~Harrison, S.~Kachru, G.~Torroba and H.~Wang,
  ``Aspects of holography for theories with hyperscaling violation,''
  arXiv:1201.1905.

\bibitem{hyper4}   K.~Narayan,
  ``On Lifshitz scaling and hyperscaling violation in string theory,''
  arXiv:1202.5935.

\bibitem{hyper5}   B.~S.~Kim,
  ``Schr\'odinger Holography with and without Hyperscaling Violation,''
  arXiv:1202.6062.

\bibitem{hsingh}   H.~Singh,
  ``Lifshitz/Schr\'odinger Dp-branes and dynamical exponents,''
  arXiv:1202.6533.

\bibitem{dey1}   P.~Dey and S.~Roy,
  ``Lifshitz-like space-time from intersecting branes in string/M theory,''
  arXiv:1203.5381.

\bibitem{dey2}   P.~Dey and S.~Roy,
  ``Intersecting D-branes and Lifshitz-like space-time,''
  arXiv:1204.4858.
    
\bibitem{swingleprl} B. Swingle, 
``Entanglement Entropy and the Fermi Surface,''
Phys. Rev. Lett. {\bf 105}, 050502 (2010). [arXiv:0908.1724]

\bibitem{kiritsis}     C.~Charmousis, B.~Gouteraux, B.~S.~Kim, E.~Kiritsis and R.~Meyer,
  ``Effective Holographic Theories for low-temperature condensed matter systems,''
  JHEP {\bf 1011}, 151 (2010).
  [arXiv:1005.4690]

\bibitem{trivedi}     N.~Iizuka, N.~Kundu, P.~Narayan and S.~P.~Trivedi,
  ``Holographic Fermi and Non-Fermi Liquids with Transitions in Dilaton Gravity,''
  JHEP {\bf 1201}, 094 (2012).
  [arXiv:1105.1162]
  
\bibitem{kiritsis2}
  B.~Gouteraux and E.~Kiritsis,
  ``Generalized Holographic Quantum Criticality at Finite Density,''
  JHEP {\bf 1112}, 036 (2011).
  [arXiv:1107.2116]
  
\bibitem{klich} D.~Gioev and I.~Klich, 
``Entanglement entropy of fermions in any dimension and the Widom conjecture,''
Phys. Rev. Lett. {\bf 96}, 100503 (2006). [arXiv:quant-ph/0504151]

\bibitem{andy}    S.~Ferrara, R.~Kallosh and A.~Strominger,
  ``N=2 extremal black holes,''
  Phys.\ Rev.\ D {\bf 52}, 5412 (1995).
  [hep-th/9508072]

\bibitem{sen}   A.~Sen,
  ``Black hole entropy function and the attractor mechanism in higher derivative gravity,''
  JHEP {\bf 0509}, 038 (2005).
  [hep-th/0506177]

\bibitem{prakash}   K.~Goldstein, S.~Kachru, S.~Prakash and S.~P.~Trivedi,
  ``Holography of Charged Dilaton Black Holes,''
  JHEP {\bf 1008}, 078 (2010).
  [arXiv:0911.3586]

\bibitem{ssfl} S.~Sachdev,   ``A model of a Fermi liquid using gauge-gravity duality,''
  Phys.\ Rev.\  D {\bf 84}, 066009 (2011).
  [arXiv:1107.5321]
  
\bibitem{lizasean} S.~A.~Hartnoll and L.~Huijse,
  ``Fractionalization of holographic Fermi surfaces,''
  arXiv:1111.2606.

\bibitem{ffl1} T.~Senthil, S.~Sachdev, and M.~Vojta, ``Fractionalized Fermi liquids,''
Phys. Rev. Lett. {\bf 90}, 216403 (2003).
[arXiv:cond-mat/0209144]
 
\bibitem{neto1} A.~H.~Castro Neto and E.~Fradkin, ``Bosonization of the Low Energy Excitations of Fermi Liquids,''
Phys. Rev. Lett. {\bf 72}, 1393 (1994). [arXiv:cond-mat/9304014]

\bibitem{neto2} A.~H.~Castro Neto and E.~Fradkin, ``Exact solution of the Landau fixed point via bosonization,''
Phys. Rev. B {\bf 51}, 4084 (1995). [arXiv:cond-mat/9310046]

\bibitem{haldane} F. D. M. Haldane, ``Luttinger's Theorem and Bosonization of the Fermi Surface,''
Proceedings of the International School of Physics ``Enrico Fermi'', Course CXXI ``Perspectives in Many-Particle Physics,'' eds. R. A. Broglia and J. R. Schrieffer (North-Holland, Amsterdam 1994). [arXiv:cond-mat/0505529]

\bibitem{marston} A. Houghton, H.-J. Kwon, and J. B. Marston, ``Multidimensional Bosonization,''
Advances in Physics {\bf 49}, 141 (2000). [arXiv:cond-mat/9810388]

\bibitem{lawler1} M.~J.~Lawler, V.~Fernandez, D.~G.~Barci, E.~Fradkin, and L.~Oxman, 
``Non-perturbative behavior of the quantum phase transition to a nematic Fermi fluid,''
Phys. Rev. B {\bf 73} , 085101 (2006). [arXiv:cond-mat/0508747]

\bibitem{lawler2} M.~J.~Lawler and E.~Fradkin, ```Local' Quantum Criticality at the Nematic Quantum Phase Transition,''
Phys. Rev. B {\bf 75}, 033304 (2007). [arXiv:cond-mat/0605203]

\bibitem{senthilbose} D.~F.~Mross and T.~Senthil, ``Decohering the Fermi liquid: A dual approach to the Mott Transition,''
Phys. Rev. B {\bf 84}, 165126 (2011). [arXiv:1107.4125]

\end{thebibliography}
\end{document}